\documentclass[journal,twoside,web]{ieeecolor}

\usepackage{generic}
\usepackage{cite}
\usepackage{amsmath,amssymb,amsfonts}
\usepackage{algorithmic}
\usepackage{graphicx}
\usepackage{algorithm,algorithmic}
\usepackage{hyperref}
\usepackage{textcomp}
\usepackage{setspace}
\usepackage{graphics} 
\usepackage{epsfig} 
\usepackage{amsmath} 
\usepackage{amssymb}  
\usepackage{cancel}

\allowdisplaybreaks 

\usepackage{hyperref}
\usepackage{graphicx}
\usepackage{subcaption}
\usepackage{ntheorem}
\newtheorem{theorem}{Theorem}
\newtheorem{rem}{Remark}
\newtheorem{defn}{Definition}
\newtheorem{lemma}{Lemma}
\newtheorem{prop}{Proposition}
\newtheorem{corollary}{Corollary}

\renewtheoremstyle{break}%
  {\item{\theorem@headerfont
          ##1\
          ##2\theorem@separator}\hskip\labelsep\relax\nobreakitem}%
  {\item{\theorem@headerfont
          ##1\ ##2\ (##3)\theorem@separator}}
\makeatother

\theoremheaderfont{\kern-0cm\normalfont\bfseries} 
\theoremstyle{break}

\usepackage{cite}

\def\BibTeX{{\rm B\kern-.05em{\sc i\kern-.025em b}\kern-.08em
    T\kern-.1667em\lower.7ex\hbox{E}\kern-.125emX}}
\begin{document}
\title{Leader-Follower Density Control of Spatial Dynamics in Large-Scale Multi-Agent Systems}
\author{Gian Carlo Maffettone, Alain Boldini, Maurizio Porfiri, Mario di Bernardo
\thanks{
This work was developed with support from the project PRIN 2022 MENTOR and of the NSF Grant CMMI-1932187. (Corresponding authors: Mario di Bernardo, Maurizio Porfiri, who contributed equally)}
\thanks{Gian Carlo Maffettone is with the Modeling and Engineering Risk and Complexity program at the Scuola Superiore Meridionale, Italy and with the Tandon School of Engineering, New York University, Brooklyn, USA (e-mail: giancarlo.maffettone@unina.it).}
\thanks{Alain Boldini is with the New York Institute of Technology,  USA (e-mail: aboldini@nyit.edu).}
\thanks{Maurizio Porfiri is with the Tandon School of Engineering, New York University, Brooklyn, USA, (e-mail: mporfiri@nyu.edu).}
\thanks{Mario di Bernardo is with the University of Naples Federico II, Naples, Italy (e-mail: mario.dibernardo@unina.it).}
}


\maketitle
\thispagestyle{empty}

\begin{abstract}
We address the problem of controlling the density of a large ensemble of follower agents by acting on a group of leader agents that interact with them. Using coupled partial integro-differential equations to describe leader and follower density dynamics, we establish feasibility conditions and develop two control architectures ensuring global stability. The first employs feed-forward control on the followers' and a feedback on the leaders' density. The second implements a dual feedback loop through a reference-governor that adapts the leaders' density based on both populations' measurements. Our methods, initially developed in a  one-dimensional setting, are extended to multi-dimensional cases, and validated through numerical simulations for representative control applications, both for groups of infinite and finite size.
\end{abstract}


\section{Introduction}
\label{sec:introduction}

\IEEEPARstart{O}{rchestrating} the collective spatial organization of multi-agent systems is crucial in fields such as traffic control \cite{siri2021freeway}, collective additive manufacturing \cite{petersen2019review}, synthetic biology \cite{del2018future}, swarm robotics \cite{Brambilla2013}, and environmental management \cite{zahugi2013oil}. Leader-follower control strategies, where leader agents steer the behavior of a group of follower agents, are widely applied in these areas and other control applications \cite{cruz1978leader, lombardi2020nonverbal, shahal2020synchronization, zienkiewicz2015leadership}.
For instance, the use of autonomous vehicles can improve traffic flows by avoiding stop-and-go waves and reducing emissions \cite{stern2018dissipation, piacentini2018traffic}, while in swarm robotics and synthetic biology, leader-follower dynamics facilitate large group management and cellular consortia regulation \cite{armbruster2017elastic, salzano2023vivo}. A critical challenge lies in establishing analytical guarantees for the achievement of desired collective tasks. This challenge is particularly evident in shepherding control problems, where a small group of controllable leader agents (or herders) must drive a larger population of follower agents toward a specified goal region. Various applications can be formulated as shepherding problems, including environmental pollutant management using robotic systems \cite{zahugi2013oil} and search and rescue operations \cite{yuan2023multi}.
Success in these applications critically depends on properly calibrating key parameters such as the ratio between leaders and followers, and their sensing capabilities \cite{ko2020asymptotic, lama2023shepherding, sebastian2022adaptive}.

To address the curse of dimensionality that is inherent to large-scale populations, microscopic models are often replaced by partial integro-differential equations  \cite{maffettone2022continuification, maffettone2023continuification, sinigaglia2022density, Elamvazhuthi2020a, maffettone2023hybrid, fornasier2014mean, gao2023lqg, zhang2024multi, nikitin2021boundary, boldini_stigmergy}.
Drawing from mixture theory \cite{mills1966incompressible},
we synthesize a macroscopic control action to steer the density of the leaders so as to indirectly control that of the followers. First, we develop a feed-forward control scheme where the follower dynamics is tamed by making the leaders density converge towards a predetermined reference. Next, using a reference-governor approach, we make such a density a function of the actual followers' density and, hence, develop a dual-feedback control strategy comprising an inner loop on the leaders' density and an outer loop on the followers'.
All our findings are corroborated by compelling numerical examples. 

\subsection{Related work and main contributions}\label{sec:related_workds}

Research in mean-field optimal control has addressed macroscopic leader-follower formulations, examining the microscopic agents' influence on infinite follower populations \cite{fornasier2014mean2, ascione2023mean}. Similar models arise in traffic control, where non-autonomous vehicles are modeled as a continuum and autonomous vehicles as dynamic constraints \cite{goatin2023interacting, lattanzio2011moving}. Analogous to our approach, \cite{bongini2017optimal} and \cite{almi2023} model the group dynamics through two coupled continuum equations representing the densities of both populations, considering both stochastic and deterministic followers. While optimal strategies exist, their explicit form and implementation through feedback control remain a pressing open question that our research addresses.


The challenge of configuring followers into predefined structures using macroscopic formulations, sometimes referred to as multi-agent deployment, is discussed in works such as \cite{wei2019pde, frihauf2010leader, zhang2024distributed, zhang2024multi}. Similarly, the problem of the deployment of followers into predefined curves is considered in \cite{khansili2023pde, wei2019multi}, and that of planar formation control is addressed in \cite{meurer2011finite}.
In these scenarios, differently from our approach, the followers' dynamics are typically represented by a heat equation, which macroscopically models a consensus protocol.

We also report significant work in deriving consensus-seeking strategies in networks of distributed parameter systems described by wave equations \cite{aguilar2020leader, fu2019containment,deutscher2021robust}.

Our main contributions are: ($i$) We propose a model that distills leader-follower spatial organization to its essentials: controllable leaders, diffusive followers, and non-reciprocal repulsive interactions. This reveals that desired collective behaviors can be achieved without requiring sophisticated interactions and communication of existing models \cite{almi2023, zhang2024multi, sebastian2022adaptive}. Our framework's effectiveness stems from its focus on basic principles--asymmetric interactions and controllable driving forces--providing a simpler yet powerful alternative to previous approaches.
($ii$) We derive analytical conditions for the existence of desired steady-state solutions for a density control problem, considering leader-follower ratios, follower characteristics, and target densities.
($iii$) We propose two control schemes with proven global stability. Differently from \cite{almi2023}, where the well-posedness of optimal control problems is considered, here we derive closed form feedback solutions, while ensuring convergence.
($iv$) We implement our solution on populations of finite size, connecting with continuification-based control \cite{maffettone2022continuification, maffettone2023continuification, maffettone2023hybrid, nikitin2021continuation}. Convergence guarantees are analytically derived in the limit of groups of infinite size. Hence, the application of our control strategy for a swarm of finite size bridges macroscopic and microscopic approaches, providing a novel feedback control solution to leader-follower problems of relevance in multiple applications \cite{siri2021freeway, petersen2019review, del2018future, Brambilla2013}.

\section{Mathematical preliminaries}\label{sec:math_prelim}
\begin{defn}[Unit circle]
    We define $\mathcal{S}:=[-\pi, \pi]$ as the unit circle.
\end{defn}
\begin{defn}[$\mathcal{L}^p$ norms on $\mathcal{S}$ \cite{axler2020}]
    Given a scalar function of $\mathcal{S}$ and time, $h:\mathcal{S}\times\mathbb{R}_{\geq0}\rightarrow\mathbb{R}$, we define its $\mathcal{L}^p$-norm on $\mathcal{S}$ as
\begin{align}
    \Vert h(\cdot, t)\Vert_p := \left( \int_{\mathcal{S}} \vert h(x, t)\vert^p \,\mathrm{d}x\right)^{1/p}.
\end{align}
For $p=\infty$,
\begin{align}
    \Vert h(\cdot, t)\Vert_\infty :=\mathrm{ess} \,\mathrm{sup}_\mathcal{S} \vert h(x, t)\vert,
\end{align}
where $\mathrm{ess} \,\mathrm{sup}$ indicates the essential supremum.
For the sake of brevity, we also denote these norms as $\Vert h\Vert_p$, without explicitly indicating their space and time dependence. 
\end{defn}
\begin{lemma} [H$\ddot{\mathrm{o}}$lder's inequality \cite{axler2020}]\label{lem:holder}
     Given $f_1, \dots, f_n \in \mathcal{L}^p(\mathcal{S})$, we have
    \begin{align}
        \left\Vert \prod_{i=1}^n f_i\right\Vert_1 \leq \prod_{i=1}^n \Vert f_i\Vert_{p_i}, \;\;\mathrm{if } \;\; \sum_{i=1}^n \frac{1}{p_i} = 1.
    \end{align}
    For instance, if $n=2$, we have $\Vert f_1f_2\Vert_1\leq \Vert f_1\Vert_2 \Vert f_2\Vert_2$, as well as $\Vert f_1f_2\Vert_1\leq \Vert f_1\Vert_1 \Vert f_2\Vert_\infty$.
\end{lemma}
We denote with \textquotedblleft{} $*$ \textquotedblright{} the convolution operator. When referring to periodic domains and functions, the operator needs to be interpreted as a circular convolution \cite{jeruchim2006simulation}.  
We identify time and space partial differentiation with the subscripts $t$ and $x$, respectively.Integrals with no explicit bounds are to be interpreted as indefinite (returning functions). Conversely, when the bounds are provided, they are definite integrals, returning constants, or, eventually, functions of the integral bounds. 
We denote by $\mathcal{C}^p(\mathcal{S})$, with $p\in\mathbb{N}_{\geq0}$, the space of functions that are differentiable $p$ times with a continuous $p$-th derivative on $\mathcal{S}$.
We use $W^{k,p}(\mathcal{S})$ for the Sobolev space of functions defined on $\mathcal{S}$ (the weak derivatives up to order $k$ are in $\mathcal{L}^p(\mathcal{S})$). Furthermore, when $p=2$, we use $H^k(\mathcal{S}) := W^{k, 2}(\mathcal{S})$ \cite{evans2022partial}.

\begin{lemma}[Poincaré-Wirtinger inequality for $\mathcal{S}$ \cite{heinonen2001lectures}]\label{lem:poincare}
Assuming $1\leq p \leq\infty$, for any function $u\in W^{1,p}(\mathcal{S})$ with null integral mean (that is, $\int_\mathcal{S}u\,\mathrm{d}x = 0$),
the following inequality holds:
\begin{align}
    \left(\Vert u \Vert_p\right)^p \leq C(p) \,\left(\Vert u_x\Vert_p\right)^p, 
\end{align}
where $C(p)>0$ is called the Poincaré constant. For $p=2$, $C = 1$  \cite{payne1960optimal, bebendorf2003note}.
\begin{rem}\label{rem:hd_Poincaré_wirt}
    The inequality can be posed for more general bounded connected open domains $\Omega\subset \mathbb{R}^d$ with Lipschitz boundary \cite{gasinski2006nonlinear, evans2022partial}. In this case, for any function $u\in W^{1,p}(\Omega)$ such that $\int_\Omega u\, \mathrm{d}\mathbf{x} = 0$, the inequality reads 
    \begin{align}
        \left(\Vert u \Vert_p\right)^p \leq C(p, \Omega) \left(\Vert \nabla u\Vert_p\right)^p, 
    \end{align}
    where $\nabla u$ is the gradient of $u$ (notice that $\Vert \nabla u \Vert_p^p = \sum_{i=1}^d \Vert u_{x_i}\Vert_p^p$, with $x_i$ being the $i$-th coordinate in $\Omega$).
\end{rem}
\begin{rem}\label{rem:poincare_wirt_const}
    When $p=2$, the optimal Poincaré constant for smooth bounded Lipschitz domains is $C = \tilde{d}/\pi$, where $\tilde{d}$ is least upper bound of the set of all distances between pairs of points in $\Omega$ \cite{payne1960optimal} (for the periodic domains we consider, $\tilde{d} = \pi$).
\end{rem}
\end{lemma}
\begin{lemma}[Comparison lemma \cite{khalil2002nonlinear}]\label{lemma:comparison_lemma}
    Given  a scalar non-autonomous ODE $v_t = f(t, v)$, with $v(t_0) =
    v_0$, where $f$ is continuous in $t$ and locally Lipschitz in $v$, if a scalar function $u$ fulfills the differential inequality
    \begin{align}
        u_t \leq f(t, u(t)), \;\;u(t_0)\leq v_0,
    \end{align}
    then
    \begin{align}
        u(t)\leq v(t),\;\;\forall\,t\geq t_0.
    \end{align}
\end{lemma}

\begin{lemma}\label{lemma:bounding_sys}
    Consider the one-dimensional dynamical system
    \begin{align}\label{eq:system_rev}
        \eta_t(t) = -\beta \eta(t) + \gamma \mathrm{exp}(-Kt)\eta(t)+\delta\mathrm{exp}(-Kt)\sqrt{\eta(t)},
    \end{align}
    with initial condition $\eta(t_0) = \eta_0\geq 0$, and parameters $\beta, K>0$ and $\gamma, \delta \geq 0$. We have that $\lim_{t\to\infty} \eta(t)=0$, for any $\eta_0$ (global asymptotic stability).
\end{lemma}
\begin{proof}
    Under the change of variable $\tilde{\eta} = \sqrt{\eta}$ and assuming $\tilde{\eta}\geq0$, we obtain
\begin{align}\label{eq:change_variables_rev}
\tilde{\eta}_t(t) = \hat{\alpha}(t) \tilde{\eta}(t) + \frac{\delta}{2} \mathrm{exp}(-Kt),
\end{align}
where $\hat{\alpha}(t) = -\frac{\beta}{2} + \frac{\gamma}{2}\mathrm{exp}(-Kt)$. Since \eqref{eq:change_variables_rev} describes a linear time-varying ODE with a time dependent forcing term, we can write \cite{rugh1996linear}
\begin{align}\label{eq:decompose}
    \tilde{\eta}(t) = \tilde{\eta}_h(t) + \tilde{\eta}_p(t),
\end{align}
where 
\begin{align}\label{eq:homog_sol}
    \tilde{\eta}_h(t) = \Phi(t, t_0) \tilde{\eta}(t_0),
\end{align}
is the solution to the homogeneous problem with
\begin{multline}
    \Phi(t, t_0) = \mathrm{exp}\left[\int_{t_0}^t \hat{\alpha}(\tau)\,\mathrm{d}\tau\right]= \mathrm{exp}\left[-\frac{\beta}{2}(t-t_0)\right]\times\\\times\mathrm{exp}\left[\frac{\gamma}{2K}\left(\mathrm{exp}(-Kt_0)-\mathrm{exp}(-Kt)\right)\right],
\end{multline}
and
\begin{align}\label{eq:particular_solution}
    \tilde{\eta}_p(t) = \frac{\delta}{2}\int_{t_0}^t \Phi(t, \sigma)\mathrm{exp}(-K\sigma)\,\mathrm{d}\sigma.
\end{align}
Since, for any $t\geq t_0$,
\begin{align}
    1\leq\mathrm{exp}\left[\frac{\gamma}{2K}\left(\mathrm{exp}(-Kt_0)-\mathrm{exp}(-Kt)\right)\right] \leq\Gamma,
\end{align}
with 
\begin{align}
    \Gamma = \mathrm{exp}\left[\frac{\gamma}{2K}\mathrm{exp}(-Kt_0)\right],
\end{align}
we have, for $t\geq t_0$,
\begin{align}\label{eq:unif_stab}
    0<\Phi(t, t_0) \leq \Gamma \mathrm{exp}\left[-\frac{\beta}{2}(t-t_0)\right],
\end{align}
proving exponential stability of \eqref{eq:change_variables_rev}. 
Exploiting the positivity of the integrands in \eqref{eq:particular_solution} and the bound \eqref{eq:unif_stab}, we also have that
\begin{multline}\label{eq:particular_sol_bound}
    0\leq\tilde{\eta}_p(t)\leq\frac{\delta\Gamma}{2}\int_{t_0}^t \mathrm{exp}\left[-\frac{\beta}{2}(t-\sigma)\right]\mathrm{exp}(-K\sigma)\,\mathrm{d}\sigma\\
    = \frac{\delta\Gamma}{2K-\beta}\bigg[\mathrm{exp}\left(\left(\frac{\beta}{2}-K\right)t_0\right)\times\\\times\mathrm{exp}\left(-\frac{\beta}{2} t\right) - \mathrm{exp}(-K t)\bigg].
\end{multline}
Once $t_0$ is fixed, for $t\to\infty$, $\tilde\eta_h$ converges to 0 for any bounded $\eta_0$ due to its exponential stability, and $\tilde\eta_p$ is bounded by a function converging to 0 (see \eqref{eq:particular_sol_bound}). This proves that also \eqref{eq:change_variables_rev} achieves the origin asymptotically, and, so does  \eqref{eq:system_rev}. Notice that, both $\tilde{\eta}_h$ and $\tilde{\eta}_p$ are non-negative for any $t\ge t_0$ (see \eqref{eq:unif_stab} and \eqref{eq:particular_sol_bound}), in agreement with our initial assumption.
\end{proof}

\section{Model} \label{sec:the_model}
We study a continuous formulation of the leader-follower control problem, where a population of leader agents (or controllers) is assigned the task of taming the behavior of a population of follower agents (or targets). 
In this framework, also adopted differently in \cite{almi2023}, two coupled equations are used to describe the spatio-temporal dynamics of the densities of the leaders and  the followers. For simplicity, differently from \cite{almi2023}, we do not consider interactions taking place between agents of the same population.
In particular, a convection-diffusion equation is used to capture the dynamics of the followers assuming that they are random walkers at the microscopic level, see e.g. \cite{lama2023shepherding, zhang2024distributed}; their interaction with the leaders being captured by a cross convection term. Conversely, the leaders' dynamics is described by a mass conservation equation influenced by some control field, say $u$, resulting in
\begin{subequations}\label{eq:themodel}
\begin{align}
    \rho_t^L (x, t) + \left[\rho^L (x, t)u(x, t)\right]_x &= 0, \label{eq:leaders}\\
    \rho_t^F (x, t) + \left[\rho^F (x, t)v^{FL}(x, t)\right]_x &= D\rho_{xx}^F(x, t), \label{eq:followers}
\end{align}
\end{subequations}
where $x\in\mathcal{S}$ and $t\in\mathbb{R}_{\geq 0}$ represent the space and time coordinates, $\rho^L, \rho^F: \mathcal{S}\times \mathbb{R}_{\geq 0}\rightarrow\mathbb{R}_{\geq 0}$ are the leaders' and followers' densities, $D\in\mathbb{R}_{\geq 0}$ weights the strength of the diffusion of the followers, and $u:\mathcal{S}\times \mathbb{R}_{\geq 0}\rightarrow\mathbb{R}$ is a velocity field to be designed in order to control the leaders' dynamics. No diffusion term is present in the leader equation, as we assume its microscopic counterpart to be deterministic.
Also,
\begin{align}\label{eq:vfl_def}
    v^{FL}(x, t) = \int_{\mathcal{S}} f({y \triangleright x})\rho^L(y, t)\,\mathrm{d}y = (f*\rho^L)(x, t)
\end{align}
is a velocity field modeling the influence of the leaders on the dynamics of the followers, where ${y \triangleright x} = (x-y+\pi) \,\mathrm{mod}(2\pi)-\pi$ is the relative position between $x$ and $y$ wrapped on $\mathcal{S}$, and $f:\mathcal{S}\rightarrow\mathbb{R}$ is a soft-core (that is, $\vert f(0)\vert < \infty$), odd interaction kernel \cite{bernoff2011primer}. To cope with the domain periodicity, we further assume $f$ to be periodic. Although the formulation is general, and any choice can be made for the kernel $f$, we fix it to be repulsive, that is,
\begin{align}\label{eq:rep_kern}
    f({x}) = \frac{\mathrm{sgn}({x})}{\mathrm{exp}\left(\frac{2\pi}{L}\right)-1} \left[\mathrm{exp}\left(\frac{2\pi-\vert {x}\vert}{L}\right) -\mathrm{exp}\left(\frac{\vert {x}\vert}{L}\right) \right],
\end{align}
where $L$ is the characteristic interaction length. 

\begin{rem}\label{rem:periodicization}
Notice that \eqref{eq:rep_kern} is the periodic version of the more standard non-periodic repulsive kernel $\hat{f}(x) = \mathrm{sgn}(x)\mathrm{exp}\left(-\frac{\vert x \vert}{L}\right)$, which is commonly studied in the literature \cite{leverentz2009asymptotic, bernoff2011primer, ko2020asymptotic} (see Appendix \ref{app:kern_periodicization}  and the Supplementary Material available at \cite{supp_material}). Due to the linearity of the convolution operator, our analysis extends to interaction kernels formed by linear combinations of terms like \eqref{eq:rep_kern}. This allows us to consider purely attractive and mixed attractive/repulsive cases. Moreover, the periodic convolution in \eqref{eq:vfl_def} is equivalent to a non-periodic convolution on $\mathbb{R}$ with a non-periodic interaction kernel \cite{jeruchim2006simulation}.
\end{rem}

By selecting $u$ in \eqref{eq:leaders} as a periodic function, such that $u(-\pi, t) = u(\pi, t)$ for all $t \in \mathbb{R}_{\geq 0}$, and imposing the periodic boundary condition
\begin{align}\label{eq:leadder_boundary}
    \rho^L(-\pi, t) = \rho^L(\pi, t), \quad \forall t \in \mathbb{R}_{\geq 0},
\end{align}
we ensure conservation of the leaders' mass $M^L$, that is, $\left(\int_\mathcal{S} \rho^L(x, t)\,\mathrm{d}x\right)_t=0$.
Equation \eqref{eq:leaders} is also complemented with its initial condition, that is
\begin{align}\label{eq:leaders_init}
    \rho^L(x, 0) = \rho^L_0(x),
\end{align}
where $\rho^L_0(x)$ is periodic and such that $\int_\mathcal{S} \rho^L_0 \,\mathrm{d}x = M^L$.

As $v^{FL}$ in \eqref{eq:followers} is periodic by construction (as it is defined as a circular convolution), the periodic boundary condition 
\begin{align}\label{eq:followers_boundary}
    \rho^F(-\pi, t) = \rho^F(\pi, t), \;\;\forall\,t\in\mathbb{R}_{\geq 0}
\end{align}
ensures the followers' mass, $M^F$, is conserved, that is $ \left(\int_\mathcal{S} \rho^F \,\mathrm{d}x\right)_t = 0$
(recalling that the derivative of a periodic function is periodic itself).
Equation \eqref{eq:followers} is complemented with its initial condition, that is
\begin{align}\label{eq:foll_init}
    \rho^F(x, 0) = \rho^F_0(x), 
\end{align}
where $\rho^F_0(x)$ is periodic and such that $\int_\mathcal{S} \rho^{F}_0 \,\mathrm{d}x = M^F$. 

\begin{rem}\label{rem:IC}
We consider $\rho_0^L, \rho_0^F \in H^2(\mathcal{S})$ (note that $\rho_0^L$ could be chosen from a less regular functional space, but we select both initial conditions from the more restrictive space). Furthermore, the periodicity of $\rho^i$, with $i=L, F$, ensures that their spatial derivatives are also periodic.
\end{rem}

We further assume that the overall mass of leaders and followers is normalized to 1, that is
\begin{align}\label{eq:tot_mass}
    M^L + M^F = 1.
\end{align}

The mathematical framework in \eqref{eq:themodel} provides a macroscopic continuum description of the shepherding control problem as described in \cite{lama2023shepherding, long2020comprehensive}, arising in applications as environmental pollutant management using robots \cite{zahugi2013oil} and search and rescue \cite{yuan2023multi}. Specifically, our model presents a mean-field approximation of the discrete shepherding dynamics analyzed in Section \ref{sec:herding} (see \eqref{eq:discrete_model}).
Furthermore, our choice of periodic domains not only simplifies derivations but also describes phenomena traditionally studied within this framework, such as traffic flow and animal behavior \cite{stern2018dissipation, Abaid2010}. Our setup can be readily adapted to general, non-periodic domains, as demonstrated in \cite{maffettone2023hybrid}.

\section{Problem statement}\label{sec:prob_statement}
We seek to find a spatially periodic control input $u$ in \eqref{eq:leaders} such that, starting from $\rho_0^F$, the leaders will displace so that the followers distribution achieves a desired configuration, that is, 
\begin{align}\label{eq:problem_statement}
        \lim_{t\to\infty} \Vert \bar{\rho}^F(\cdot) - \rho^F(\cdot, t) \Vert_2 = 0,
\end{align}
where $\bar{\rho}^F: \mathcal{S} \to \mathbb{R}_{>0}$ is the desired stationary periodic density profile for the followers. We note that, by designing $u$, we are indirectly controlling the dynamics of the followers' population by driving the density of the leaders, $\rho^L$, which, in turn, influences the followers' population through the interaction kernel $f$. 

\begin{rem}
    Notice that, in the absence of leaders, \eqref{eq:followers} describes Brownian motion of the followers at the macroscopic level.  Such a behavior represents an effective evasive strategy, as shown in \cite{zhang2024distributed}, within the context of a shepherding problem (see \cite{long2020comprehensive} for further details).
\end{rem}

\section{Feasibility analysis} \label{sec:feasibility}
\begin{figure*}[t]
     \centering
     \begin{subfigure}[b]{0.32\textwidth}
         \centering
         \includegraphics[width=\textwidth]{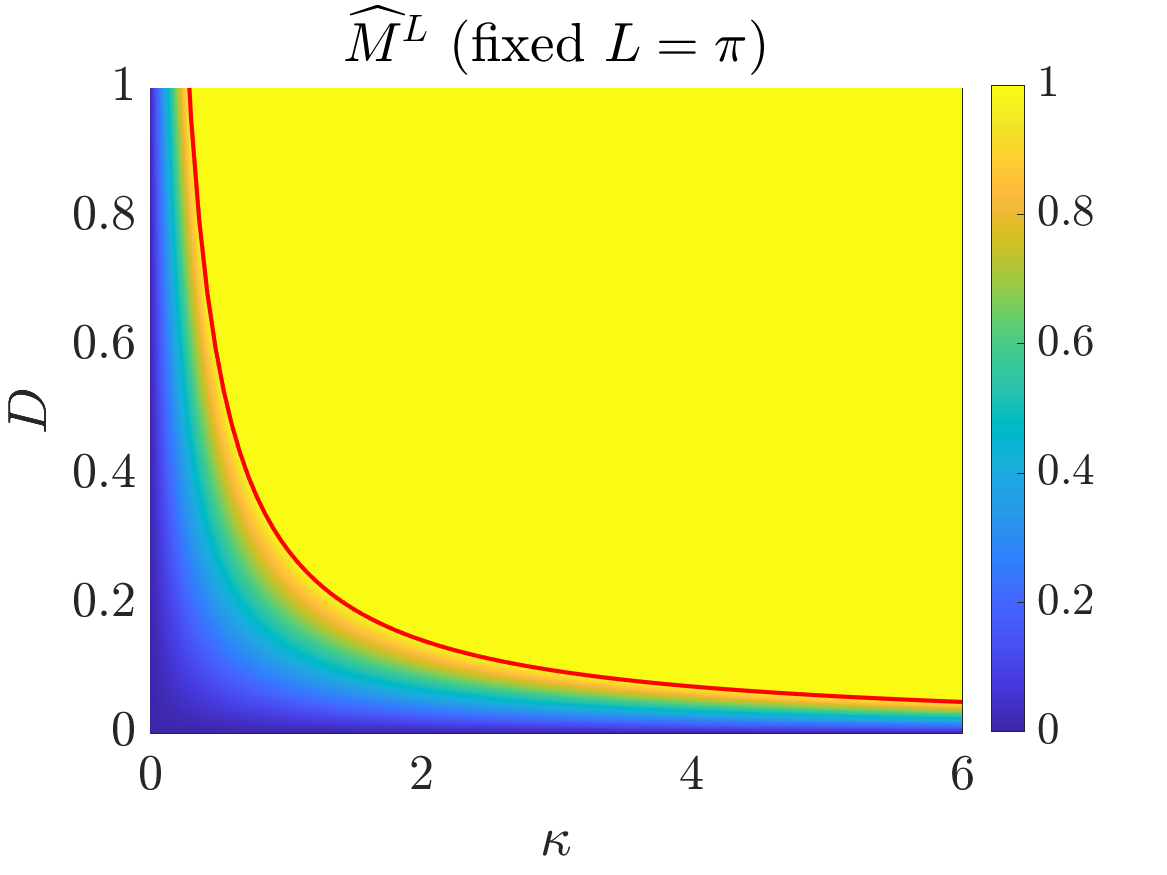}
         \caption{}
         \label{subfig::fixedL}
     \end{subfigure}
     \begin{subfigure}[b]{0.32\textwidth}
         \centering
         \includegraphics[width=\textwidth]{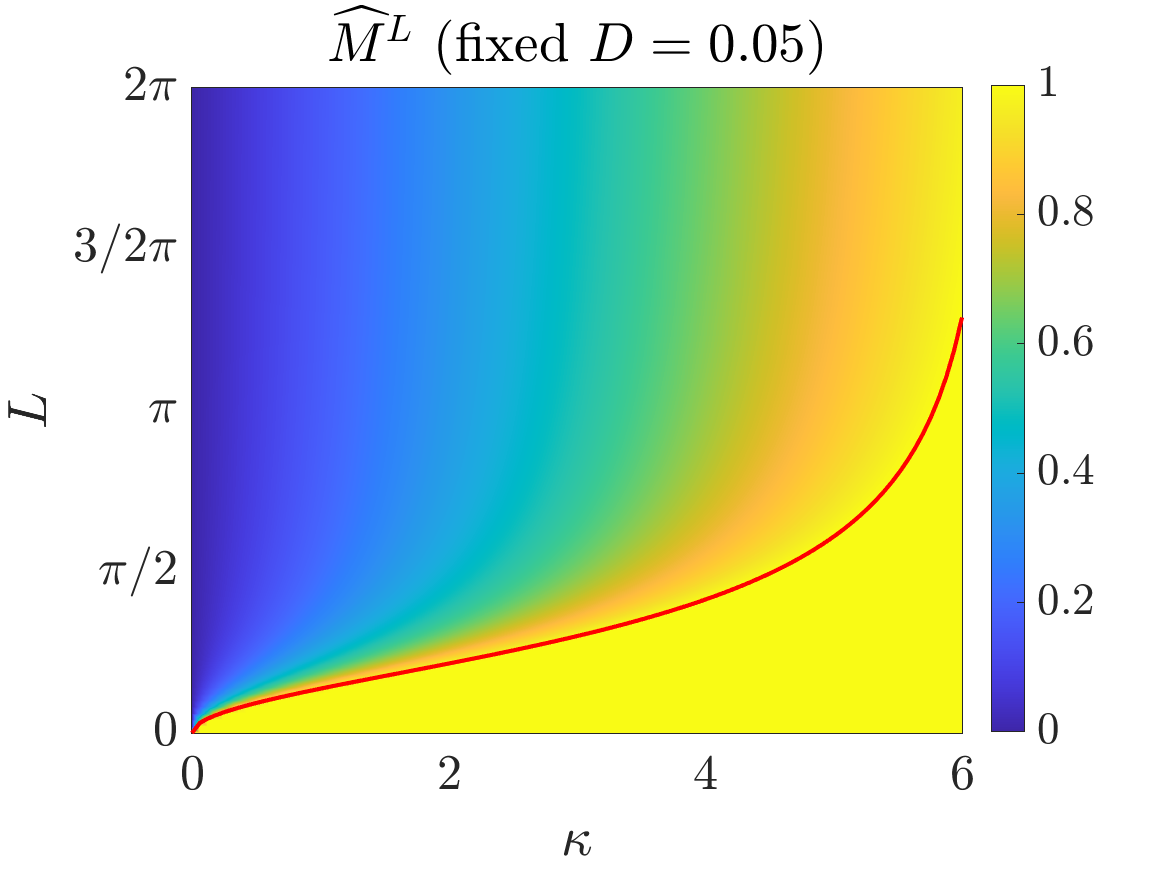}
         \caption{}
         \label{subfig::fixedD}
     \end{subfigure}
     \begin{subfigure}[b]{0.32\textwidth}
         \centering
         \includegraphics[width=\textwidth]{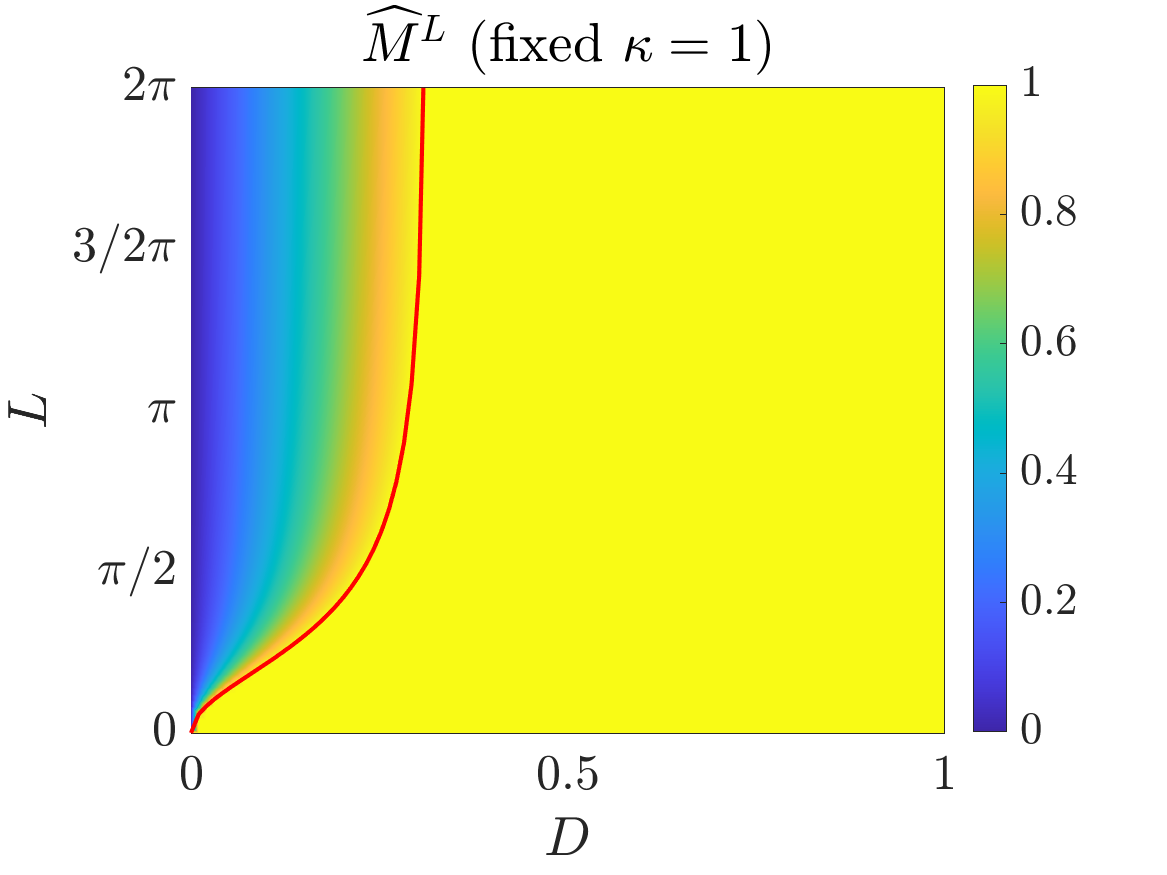}
         \caption{}
         \label{subfig::fixedk}
     \end{subfigure}
        \caption{\footnotesize Feasibility plots: minimum amount of leaders' mass, $\widehat{M}^L$ for (a) fixed $L$ and varying $\kappa$ and $D$, (b) fixed $D$ and varying $\kappa$ and $L$, and (c) fixed $\kappa$ and varying $D$ and $L$. In red we show the curve indicating when $\widehat{M}^L$ becomes greater than 1. $\widehat{M}^L$ has been saturated to 1 for visualization purposes.}
        \label{fig::feasibility_example}
\end{figure*}
\begin{defn}\label{def1}
We say that problem \eqref{eq:themodel}-\eqref{eq:problem_statement} admits a feasible steady-state solution (or, equivalently, that it is \textit{feasible}) if, given a followers' mass $0<M^F<1$, there exists a leaders' density $\bar{\rho}^L(x)$, fulfilling the following two conditions:
\begin{align}
        &\mathrm{1.}\;\; \bar{\rho}^L(x) \geq 0, \;\;\forall \, x \in \mathcal{S},\tag{D1.1}\label{cond_11}\\
        &\mathrm{2.}\;\; \int_{\mathcal{S}}\bar{\rho}^L(x)\,\mathrm{d}x  = M^L = 1 - M^F\tag{D1.2}\label{cond_12},
 \end{align}
and such that the desired followers' density $\bar{\rho}^F$ is a steady-state solution of \eqref{eq:followers}, upon setting $v^{FL}(x, t)= \bar{v}^{FL}(x) := (f * \bar{\rho}^L)(x)$.
\end{defn}
\begin{rem}\label{rem:lead_sol}
   We do not explicitly require $\bar{\rho}^L$ to be a solution of \eqref{eq:leaders}, as this can be ensured through an appropriate choice of the control input $u$, as demonstrated in Section \ref{sec:lead}.
\end{rem}

Hence, for the problem to be feasible, a necessary condition is for $\bar{\rho}^F$ to be a steady-state solution of \eqref{eq:followers}
\begin{align}\label{eq:eq13}
    \left[\bar{\rho}^F (x)\bar{v}^{FL}(x)\right]_x = D\bar{\rho}_{xx}^F(x).
\end{align}
By spatial integration, recalling $\bar{\rho}^F(x)\neq 0$ $\forall \,x\in\mathcal{S}$ (see Section \ref{sec:prob_statement}), we can recover the expression of the velocity field $\bar{v}^{FL}$ such that, being the problem set in $H^2(\mathcal{S})$, \eqref{eq:eq13} is fulfilled almost everywhere,
\begin{align}\label{eq:vfl_A}
    \bar{v}^{FL}(x) = D \frac{\bar{\rho}^F_x(x)}{\bar{\rho}^F(x)} + \frac{A}{\bar{\rho}^F(x)},
\end{align}
where $A$ is an arbitrary integration constant. Note that, by reformulating the problem in $\mathcal{C}^2(\mathcal{S})$ and not in $H^2(\mathcal{S})$, \eqref{eq:vfl_A} would hold point wise.

To find $A$, we notice that, as $f$ is odd, from Fubini's theorem for convolutions \cite{royden1968real}, we must have
\begin{multline}\label{eq:Fubini}
    \int_{\mathcal{S}} \bar{v}^{FL}(x) \,\mathrm{d}x = \int_{\mathcal{S}} (f*\bar{\rho}^L)(x) \,\mathrm{d}x = \\=\int_{\mathcal{S}} f(x) \,\mathrm{d}x \int_{\mathcal{S}} \bar{\rho}^L(x) \,\mathrm{d}x = 0.
\end{multline}
Then, using \eqref{eq:Fubini}, from \eqref{eq:vfl_A} we can derive
\begin{align}
    A = -\frac{D \int_{\mathcal{S}}{\bar{\rho}_x^F(x)/\bar{\rho}^F(x)} \,\mathrm{d}x}{\int_\mathcal{S} 1/\bar{\rho}^F(x) \,\mathrm{d}x}=  -\frac{ D \left[\mathrm{log}(\bar{\rho}^F(x))\right]_{-\pi}^\pi}{\int_\mathcal{S} 1/\bar{\rho}^F(x) \,\mathrm{d}x} = 0,
\end{align}
because of the periodicity of $\bar{\rho}^F$. Thus, setting $A=0$ in \eqref{eq:vfl_A} we find
\begin{align} \label{eq:vfl_bar}
    \bar{v}^{FL}(x) = D \frac{\bar{\rho}^F_x(x)}{\bar{\rho}^F(x)}.
\end{align}

Given the expression of $\bar{v}^{FL}$ in \eqref{eq:vfl_bar}, knowing that $\bar{v}^{FL} = f*\bar{\rho}^L$, and assuming the expression of the repulsive kernel in \eqref{eq:rep_kern}, we can recover the reference leaders' density $\bar{\rho}^L$ by deconvolution \cite{wing1991primer} (see Appendix \ref{app:deconv} for more details), yielding
\begin{align}\label{eq:rho_l_bar}
    \bar{\rho}^L(x) = \frac{\bar{v}^{FL}_x(x)}{2} - \frac{1}{2L^2}\int \bar{v}^{FL}(x) \,\mathrm{d}x + B,
\end{align}
where $B$ is an arbitrary constant. The deconvolution operation does not automatically guarantee  that the resulting leaders' density $\bar{\rho}^L$ is feasible according to Definition \ref{def1}. Then, problem \eqref{eq:themodel}-\eqref{eq:problem_statement} is feasible if there exists a constant $B$ in \eqref{eq:rho_l_bar} such that conditions \eqref{cond_11} and \eqref{cond_12} in Definition \ref{def1} hold.

Using \eqref{eq:rho_l_bar}, \eqref{cond_11}, and \eqref{cond_12}, we can derive a lower bound on the mass of leaders needed to make the problem feasible as a function of the kernel parameters, the diffusivity of the followers, and the desired density profile. In what follows, we normalize the desired followers' density as
\begin{align}\label{eq:norm_rho_f_des}
    \bar{\rho}^F(x) = M^F \hat{\rho}^F (x),
\end{align}
with
\begin{align}
    \int_{\mathcal{S}} \hat{\rho}^F(x) \,\mathrm{d}x = 1.
\end{align}
\begin{theorem}\label{proposition:prop1}
   The problem described by \eqref{eq:themodel} and \eqref{eq:problem_statement} with $f$ from \eqref{eq:rep_kern} is feasible according to Definition \ref{def1} if and only if, given $\hat{\rho}^F$ as in \eqref{eq:norm_rho_f_des}, the leaders' mass $M^L$ is such that
    \begin{align}\label{cond:feas_2}
\widehat{M}^L\leq M^L<1
    \end{align}
with 
$$
\widehat{M}^L =\max_x\left\{h(x)\right\},
$$
    where
    \begin{align}\label{eq:h_def}
        h(x) = -\pi D g_1(x) + \frac{\pi D}{L^2} g_2(x) - \frac{DC}{2L^2},
    \end{align}
    and
    \begin{align}
       g_1(x) &=\left[ \log (\hat{\rho}^F(x))\right]_{xx} = \left(\frac{\hat{\rho}^F_x(x)}{\hat{\rho}^F(x)}\right)_x,\label{eq:g1}\\
        g_2(x) &= \log (\hat{\rho}^F(x)),\label{eq:g2}\\
        C &=  \int_\mathcal{S}{\log (\hat{\rho}^F(x))}\,\mathrm{d}x.\label{eq:C}
    \end{align}
\end{theorem}
\begin{proof}
We first prove sufficiency ($\Rightarrow$). Given the expression for $\bar\rho^L$ in \eqref{eq:rho_l_bar}, we seek for conditions under which it is feasible according to Def. \ref{def1}.
Substituting \eqref{eq:norm_rho_f_des} into \eqref{eq:vfl_bar} we rewrite $\bar{v}^{FL}$ as
\begin{align}\label{eq:vlfbarhat}
    \bar{v}^{FL}(x) = D\frac{\hat{\rho}^F_x(x)}{\hat{\rho}^F(x)}.
\end{align}
Using this expression for $\bar{v}^{FL}$ in the definition of $\bar\rho^L$ in \eqref{eq:rho_l_bar}, we obtain
    \begin{align}
    \int_{\mathcal{S}} \bar{\rho}^L(x) \,\mathrm{d}x 
    = -\frac{D\,C}{2L^2} +2\pi B, 
\end{align}
where we used the periodicity of $\bar{v}^{FL}$ and fixed $C$ as in \eqref{eq:C}.
Now, to fulfill \eqref{cond_12} in Definition \ref{def1}, we select the arbitrary constant $B$ in \eqref{eq:rho_l_bar} as
\begin{align} 
    B = \frac{1}{2\pi}\left(1 - M^F +\frac{D\,C}{2L^2}\right).
\label{eq:B}
\end{align}
Substituting this expression of $B$ into \eqref{eq:rho_l_bar}, and computing $\bar{v}^{FL}_x$,  from \eqref{cond_11} we have
\begin{align}\label{eq:rho_l_des}
    \bar{\rho}^L(x) = \frac{D}{2} g_1(x) - \frac{D}{2L^2} g_2(x)+ \frac{1}{2\pi}\left(1 - M^F +\frac{D\,C}{2L^2}\right)\geq 0,
\end{align}
with $g_1$ and $g_2$ given by \eqref{eq:g1} and \eqref{eq:g2}, respectively.

Hence, problem \eqref{eq:problem_statement} admits a feasible solution if \eqref{eq:rho_l_des} is fulfilled. From \eqref{eq:rho_l_des}, knowing that $M^L = 1 - M^F$, it follows
\begin{align}\label{eq:ML_cond}
    M^L \geq -\pi D g_1(x) + \frac{\pi D}{L^2} g_2(x) - \frac{DC}{2L^2}  = h(x),\;\;\forall\,x\in\mathcal{S},
\end{align}
which is always fulfilled under \eqref{cond:feas_2}, thus proving sufficiency.

To prove necessity ($\Leftarrow$), we assume feasibility, that is, we know there exists some non-negative $\bar{\rho}^L$ summing to $M^L\in(0, 1)$, making $\bar{\rho}^F$ a steady-state solution of \eqref{eq:followers}. Hence, the steps from \eqref{eq:vlfbarhat} to \eqref{eq:ML_cond} hold by assumption. Being $M^L$ constant, it must be $M^L\geq\widehat{M}^L$ for \eqref{eq:ML_cond} to hold.
\end{proof}
\begin{rem}
    The use of Theorem \ref{proposition:prop1} (specifically, condition \eqref{cond:feas_2}) is twofold. ($i$) Given the normalized desired followers' density $\hat{\rho}^F$, one can derive a condition on the minimum amount of leaders' mass $M^L$ that makes the problem feasible. ($ii$) Given the available mass of leaders, one can identify what desired densities of followers can be effectively achieved. 
\end{rem}

\subsection{Example}\label{subsec:example}
Let us assume that the normalized desired followers' density is the von Mises distribution
\begin{align}\label{eq:vonmises}
    \hat{\rho}^F(x) = \frac{{\mathrm{exp}\left(\kappa \cos(x-\mu)\right)}}{2\pi I_0(\kappa)},
\end{align}
where $\kappa$ is the concentration coefficient, $\mu$ is the mean and $I_0$ is the modified Bessel function of the first kind of order 0. Without any loss of generality, we fix $\mu = 0$ and use \eqref{eq:g1}, \eqref{eq:g2}, \eqref{eq:C} to compute 
\begin{subequations}
\begin{align}
    g_1(x) &= -\kappa\cos(x),\\
    g_2(x) &= \kappa \cos(x) - \log[2\pi I_0(\kappa)],\\
    C &= -2\pi\log[2\pi I_0(\kappa)].
\end{align}
\end{subequations}
Substituting into \eqref{eq:h_def}, we obtain
\begin{align}
    h(x) = \pi D \left(1+\frac{1}{L^2}\right)\kappa\cos(x), 
\end{align}
whose maximum in $\mathcal{S}$ is
\begin{align}
    \widehat M^L=\max_{x\in\mathcal{S}}{h(x)} =  \pi D \left(1+\frac{1}{L^2}\right)\kappa.
\end{align}
Therefore, from Theorem \ref{proposition:prop1} the problem is feasible if 
\begin{align}\label{eq:Mlhat}
    \pi D \left(1 + \frac{1}{L^2}\right)\kappa<M^L<1
\end{align}
In Fig. \ref{fig::feasibility_example}, we report $\widehat{M}^L$ as a function of $D$, $L$ and $\kappa$ in three different scenarios. Specifically, in Fig. \ref{subfig::fixedL}, we consider $L=\pi$ and we let $\kappa$ and $D$ vary; in Fig. \ref{subfig::fixedD}, we fix $D = 0.05$ and we let $\kappa$ and $L$ vary; in Fig. \ref{subfig::fixedk}, we fix $\kappa = 1$ and let $D$ and $L$ vary. From Fig. \ref{fig::feasibility_example} we notice that a larger leaders' mass is needed for larger values of $D$ and $\kappa$, and for smaller values of $L$. This suggests that highly diffusive followers (large $D$) require more leaders for effective control, supporting the use of random walks as an evasive strategy \cite{zhang2024distributed}. Moreover, achieving more concentrated desired density profiles (large $\kappa$) demands a greater mass of leaders. Additionally, as expected, a broader area of influence (large $L$) simplifies the leaders' task.

For completeness,  we also report the resulting expression of the reference leaders' density as computed from \eqref{eq:rho_l_bar} with $B$ from \eqref{eq:B}, that is,
\begin{align}\label{eq:leaders_des_density}
    \bar{\rho}^L(x) = - \frac{D\,\kappa}{2}\left(1+\frac{1}{L^2}\right)\cos(x) + \frac{M^L}{2\pi}.
\end{align}

\section{Followers' Feed-Forward and Leaders' Feedback Control}\label{sec:control design}
\begin{figure}
    \centering
    \includegraphics[width=0.5\textwidth]{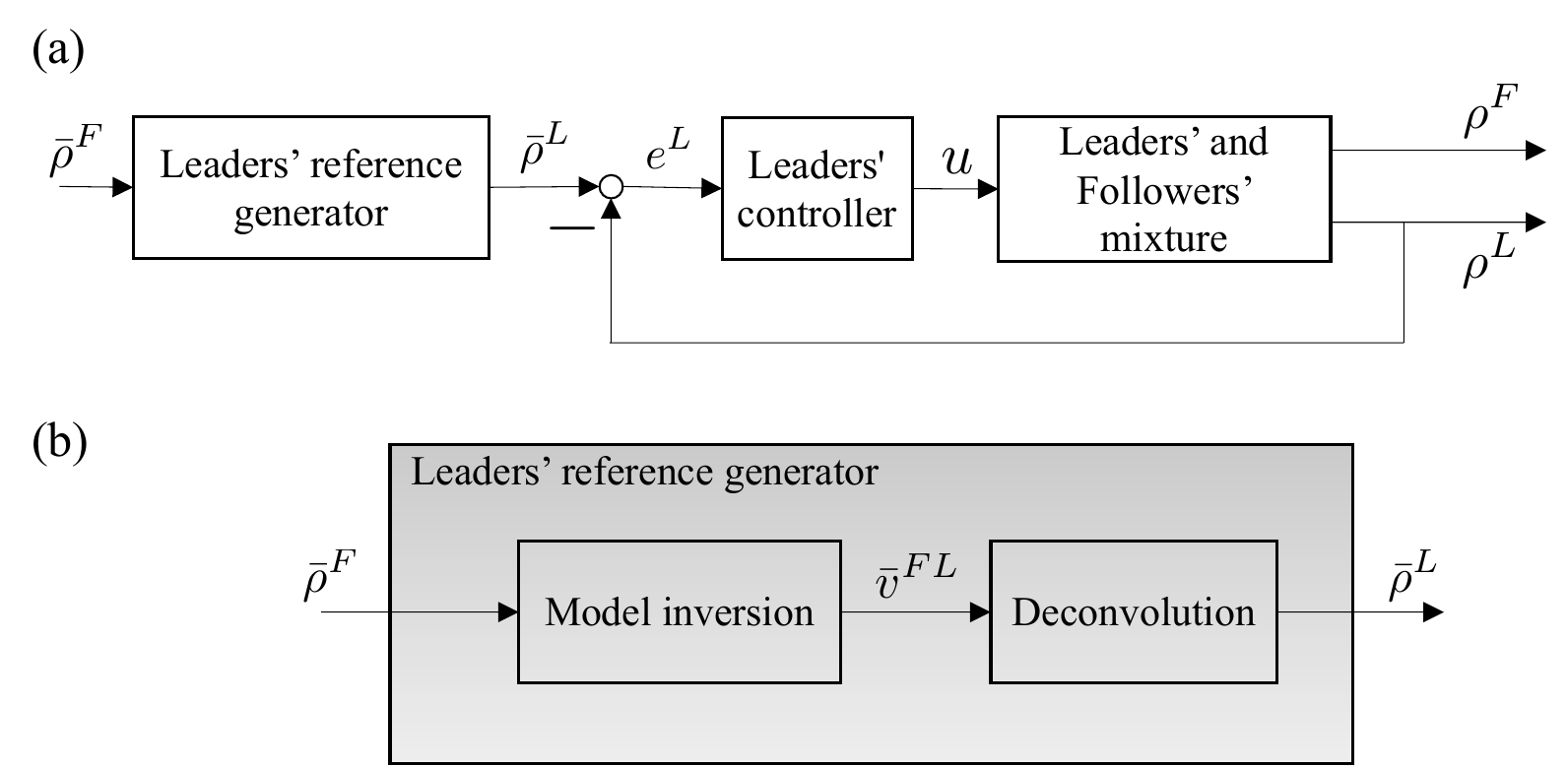}
    \caption{\footnotesize (a) Feed-forward control scheme. (b) Detail of the leaders' reference generator block (where the feasibility analysis is performed).}
    \label{fig::active_stigmergycontrol_scheme}
\end{figure}

Assuming the problem is feasible according to Definition \ref{def1}, we start by seeking an expression for $u$ in \eqref{eq:leaders} that drives the leaders' density from $\rho^L_0$ towards $\bar{\rho}^L$ (computed from \eqref{eq:rho_l_bar}, fixing $B$ as in \eqref{eq:B}) and, under appropriate conditions, renders $\bar{\rho}^F$ an asymptotically stable solution of \eqref{eq:followers}. The overall control scheme is reported in Fig. \ref{fig::active_stigmergycontrol_scheme}. 
This control solution does not use information about the followers' density $\rho^F$ when controlling the leaders, making it a feed-forward scheme with respect to the followers' dynamics. To improve robustness, we will propose a dual feedback reference-governor scheme in Section \ref{sec:ref_gov}, combining feedback on both $\rho^L$ and $\rho^F$.

\subsection{Leaders' control design}\label{sec:lead}
Given a desired density profile, $\bar{\rho}^L$, fulfilling \eqref{cond_11} and \eqref{cond_12} in Definition \ref{def1}, we want to choose $u$ in \eqref{eq:leaders} so as to drive $\rho^L$ to it. We recall that a spatially periodic $u$ ensures leaders' mass conservation (see Section \ref{sec:the_model}). 


We define the leaders' density error as
\begin{align}\label{eq:lead_err}
    e^L(x, t) =\bar{\rho}^L(x) - \rho^L(x, t).
\end{align}
Notice that the following integral condition is fulfilled:
\begin{multline}\label{eq:int_cond_leaders}
    \begin{aligned}
        \int_{\mathcal{S}} e^L(x, t) \,\mathrm{d}x = \int_{\mathcal{S}} \left(\bar{\rho}^L(x) - \rho^L(x, t)\right) \,\mathrm{d}x =\\= M^L - M^L = 0, \;\;\forall\,t\in\mathbb{R}_{\geq 0}.
    \end{aligned}
\end{multline}

\begin{theorem} [Leaders' global exponential convergence]\label{th:leaders_convergnce}
    Choosing $u$ from the spatial integration of
    \begin{align}\label{eq:lead_flux}
        \left[\rho^L(x, t)u(x, t)\right]_x = - K_L e^L(x, t),
    \end{align}
    with $K_L >0$, makes the leaders' error dynamics globally point-wisely, exponentially convergent to zero.
\end{theorem}
\begin{proof}
    From \eqref{eq:lead_err} and \eqref{eq:leaders}, the leaders' error dynamics obeys
    \begin{align}\label{eq:err_lead_interm}
        e^L_t(x, t) =  - \rho^L_t(x, t) =  \left[\rho^L(x, t)u(x, t)\right]_x,
    \end{align}
    with its initial and periodic boundary condition
    \begin{subequations}
    \begin{align}
        e^L(x, 0) &= \bar{\rho}^L(x) - \rho^L_0(x)   \;\;\forall \,x\in\mathcal{S}\\
        e^L(-\pi, t) &= e^L(\pi, t) \;\;\forall \,t\in\mathbb{R}_{\geq 0}.
    \end{align}
    \end{subequations}
    Substituting \eqref{eq:lead_flux} into \eqref{eq:err_lead_interm} yields
    \begin{align}\label{eq:linear_err_lead}
        e^L_t(x, t) = - K_L e^L(x, t),
    \end{align}
    which is linear and globally convergent. Therefore, 
    \begin{align}\label{eq:leaders_behavior}
        e^L(x, t) = e^L(x, 0)\,\mathrm{exp}\left(-K_L t\right).
    \end{align}
\end{proof}

\begin{rem}\label{rem:leaders_rem}
    The control action is obtained as described in \cite{maffettone2022continuification, maffettone2023hybrid} (neglecting interactions between agents within the same population). The control input $u$ can be derived through spatial integration of \eqref{eq:lead_flux}
\begin{align}
u(x, t) = -\frac{K_L}{\bar{\rho}^L(x)-e^L(x, t)}\int e^L(x, t),\mathrm{d}x.
\end{align}
Note that $u$ is well defined only when $\rho^L(x, t) \neq 0$ (recall that $\rho^L = \bar{\rho}^L-e^L$), since control cannot be exerted otherwise.
\end{rem}

The periodicity of $u$, which is proved next, ensures  that leaders' mass is conserved ( $\int_\mathcal{S} \rho^L_t(x, t) \,\mathrm{d}x = 0$).

\begin{corollary}\label{th:u_periodicity}
    The field $u$ obtained by spatially integrating \eqref{eq:lead_flux} is periodic, that is $u(-\pi, t) = u(\pi, t)$ $\forall\,t\in\mathbb{R}_{\geq 0}$.
\end{corollary}
\begin{proof}
    By spatially integrating \eqref{eq:lead_flux} in $\mathcal{S}$, we get 
    \begin{align}\label{eq:cor_mid}
        \int_\mathcal{S} \left[\rho^L(x, t) u(x, t)\right]_x\,\mathrm{d}x = -\int_\mathcal{S} K_Le^L(x, t) \,\mathrm{d}x = 0, 
    \end{align}
    where we used \eqref{eq:int_cond_leaders}. Expanding the first member of \eqref{eq:cor_mid} implies  
    \begin{align}\label{eq:flux_lead_periodicity}
        \rho^L(-\pi, t) u(-\pi, t) = \rho^L(\pi, t)u(\pi, t).
    \end{align}
   As $\rho^L(\pi, t) = \rho^L(-\pi, t)$ from the boundary conditions of \eqref{eq:leaders}, the thesis follows.
\end{proof}
\begin{rem}\label{rem:leaders_tracking}
    Notice that the overall control strategy for the leaders' density can be easily adapted to tracking scenarios. Indeed, if $\tilde{\rho}^L(x, t)$ is a time varying periodic density fulfilling some mass conservation principle (that is, $\int_{\mathcal{S}} \tilde{\rho}^L_t(x, t)\,\mathrm{d}x = 0$), choosing
    \begin{align}\label{eq:flux_tv}
         \left[\rho^L(x, t)u(x, t)\right]_x = - \tilde{\rho}^L_t(x, t) - K_L e^L(x, t),
    \end{align}
    still allows the error dynamics to be recast as in \eqref{eq:linear_err_lead} (in this context the error is $e^L = \tilde{\rho}^L - \rho^L$). Such a choice is also associated with a periodic velocity field $u$. In fact,
    \begin{align}\label{eq:flux_tv2}
        \int_\mathcal{S}\left[-\tilde{\rho}^L_t(x, t) - K_Le^L(x, t)\right] \,\mathrm{d}x = 0, \;\;\forall \,t\geq 0,
    \end{align}
    so that \eqref{eq:flux_lead_periodicity} and Corollary \ref{th:u_periodicity} still hold.
\end{rem}

\subsection{Followers' stability analysis}\label{subsec:active_stigmergy_stability}
Under the control action discussed in Section \ref{sec:lead}, we know leaders' density exponentially converges to $\bar{\rho}^L$.
Here, we prove that, under suitable conditions, global convergence of the followers' density towards $\bar{\rho}^F$ is also attained.

We define the followers' error as
\begin{align}
    e^F(x, t) = \bar{\rho}^F(x) - \rho^F(x, t).
\end{align}
Notice that, by construction, $\int_\mathcal{S}e^F(x, t) \,\mathrm{d}x = 0$ $\forall \, t\geq 0$.
The error dynamics is given by
\begin{multline}\label{eq:err_dyn_act_stig}
    e^F_t(x, t) = \left[\left(\bar{\rho}^F(x)-e^F(x, t)\right) v^{FL}(x, t)\right]_x \\+D \left(e_{xx}^F(x, t)-\bar{\rho}^F_{xx}(x)\right),
\end{multline}
subject to initial and periodic boundary conditions
\begin{subequations}\label{eq:IC-PBC_err}
     \begin{align}
        e^F(x, 0) &= \bar{\rho}^F(x) - \rho^F(x, 0)   \;\;\forall \,x\in\mathcal{S}\label{eq:errfoll_ic}\\
        e^F(-\pi, t) &= e^F(\pi, t) \;\;\forall \,t\in\mathbb{R}_{\geq 0}\label{eq:errfoll_PBC}.
    \end{align}
\end{subequations}

\begin{theorem}[Followers' global stability]\label{th_act_stig_stab}
    If the feasibility condition in Theorem \ref{proposition:prop1} is satisfied, and $\Vert g_1(\cdot) \Vert_\infty < 2$
    (see \eqref{eq:g1}), then the error dynamics \eqref{eq:err_dyn_act_stig} converge globally to 0 in $\mathcal{L}^2(\mathcal{S})$. 
\end{theorem}
\begin{proof}
From \eqref{eq:leaders_behavior}, we know that
    \begin{align}
        \rho^L(x, t) = \bar{\rho}^L(x) + \Phi(x, t),
    \end{align}
    where 
    \begin{align}\label{phi_reply}
        \Phi(x, t) = - \left[\bar{\rho}^L(x) - \rho^L(x, 0)\right]\mathrm{exp}(-K_L t)
    \end{align}
    represents the transient leaders' behavior. Since $v^{FL} = f*\rho^L$, \eqref{eq:err_dyn_act_stig} becomes
    \begin{multline}\label{eq:err_reply}
        e^F_t(x, t) = D \left(e_{xx}^F(x, t)-\bar{\rho}^F_{xx}(x)\right)\\+\left[\left(\bar{\rho}^F(x)-e^F(x, t)\right) (f*\bar{\rho}^L)(x)\right]_x \\    +\left[\left(\bar{\rho}^F(x)-e^F(x, t)\right) (f*\Phi)(x, t)\right]_x.
    \end{multline}
Substituting \eqref{phi_reply} into \eqref{eq:err_reply}, and recalling that, upon the fulfillment of the feasibility condition, $f*\bar{\rho}^L = \bar{v}^{FL}$, we recover
\begin{multline}\label{eq:perturbed_err}
e^F_t(x, t) = De^F_{xx}(x, t) \\- D \left[1-\mathrm{exp}(-K_L t)\right]\left[e^F(x, t)\frac{\bar{\rho}^F_x(x)}{\bar{\rho}^F(x)}\right]_x\\+
\mathrm{exp}(-K_L t)\left[(\bar{\rho}^F(x)-e^F(x, t)) (f*\rho^L_0)(x) -D\bar{\rho}^F_{x}(x)\right]_x.
\end{multline}
Choosing $\Vert e^F \Vert_2^2$ as a Lyapunov functional and recalling that $\left(\Vert e^F \Vert_2^2\right)_t = \int_\mathcal{S}e^Fe^F_t\,\mathrm{d}x$, we obtain
\begin{multline}\label{eq:Vdot_reply}
    \left(\Vert e^F(\cdot, t) \Vert_2^2\right)_t 
    = 2 D \int_\mathcal{S} e^F(x, t) e^F_{xx}(x, t) \,\mathrm{d}x +\\- 2 D \left[1-\mathrm{exp}(-K_L t)\right] \int_\mathcal{S} e^F(x, t)\left[e^F(x, t) \frac{\bar{\rho}^F_x(x, t)}{\bar{\rho}^F(x)}\right]_x  \,\mathrm{d}x \\
    +2\mathrm{exp}(-K_L t)\times\\ \int_{\mathcal{S}} e^F(x, t) \left[(\bar{\rho}^F(x)-e^F(x, t)) (f*\rho^L_0)(x)-D\bar{\rho}^F_{x}(x)\right]_x\,\mathrm{d}x,
\end{multline}
where we used \eqref{eq:perturbed_err}. Integrating by parts the terms on the right hand-side of \eqref{eq:Vdot_reply} accounting for their periodicity, and recalling that $[(e^F)^2]_x = 2e^Fe^F_x$, we establish the following identities:
\begin{subequations}\label{eq:identities}
\begin{multline}
    2 D \int_\mathcal{S} e^F(x, t) e^F_{xx}(x, t) \,\mathrm{d}x = -2D \int_\mathcal{S} (e^F_x(x, t))^2 \,\mathrm{d}x =\\= -2D\Vert e^F_x(\cdot, t) \Vert_2^2,
\end{multline}
\begin{multline}\label{eq:identity_b}
    - 2 D \left[1-\mathrm{exp}(-K_L t)\right] \int_\mathcal{S} e^F(x, t)\left[e^F(x, t) \frac{\bar{\rho}^F_x(x, t)}{\bar{\rho}^F(x)}\right]_x  \,\mathrm{d}x =\\= D\left[1-\mathrm{exp}(-K_L t)\right] \int_\mathcal{S} 2e^F_x(x, t)e^F(x, t) \frac{\bar{\rho}^F_x(x, t)}{\bar{\rho}^F(x)} \,\mathrm{d}x =\\= D\left[1-\mathrm{exp}(-K_L t)\right] \int_\mathcal{S} \left[(e^F(x, t))^2\right]_x \frac{\bar{\rho}^F_x(x, t)}{\bar{\rho}^F(x)} \,\mathrm{d}x=\\=
    -D\left[1-\mathrm{exp}(-K_L t)\right] \int_\mathcal{S} (e^F(x, t))^2 g_1(x) \,\mathrm{d}x,
\end{multline}
\begin{multline}
    -2\mathrm{exp}(-K_L t)\int_\mathcal{S} e^F(x, t) \left[e^F(x, t)(f*\rho_0^L)(x)\right]_x\,\mathrm{d}x =\\= \mathrm{exp}(-K_L t)\int_\mathcal{S} 2 e^F(x, t) e^F_x(x, t)(f*\rho_0^L)(x)\,\mathrm{d}x =\\= \mathrm{exp}(-K_L t)\int_\mathcal{S}  \left[(e^F(x, t))^2\right]_x (f*\rho_0^L)(x)\,\mathrm{d}x 
    =\\=-\mathrm{exp}(-K_L t)\int_\mathcal{S}(e^F(x, t))^2 h_1(x)\,\mathrm{d}x,
\end{multline}
\end{subequations}
with $h_1(x) = \left[(f*\rho_0^L)(x)\right]_x$.
By substituting the identities \eqref{eq:identities} into \eqref{eq:Vdot_reply}, we find
\begin{multline}\label{eq:polished_vdot}
     \left(\Vert e^F(\cdot, t) \Vert_2^2\right)_t = -2D\Vert e^F_x(\cdot, t) \Vert_2^2 \\-D\left[1-\mathrm{exp}(-K_L t)\right] \int_\mathcal{S} (e^F(x, t))^2 g_1(x) \,\mathrm{d}x \\-\mathrm{exp}(-K_L t)\int_\mathcal{S}(e^F(x, t))^2 h_1(x)\,\mathrm{d}x \\+ 2\mathrm{exp}(-K_L t)\int_\mathcal{S} e^F(x, t) h_2(x)\,\mathrm{d}x,
\end{multline}
where we posed $h_2 = [\bar{\rho}^F(f*\rho_0^L) - D \bar{\rho}^F_x]_x$. Using Poincaré-Wirtinger inequality and H$\ddot{\mathrm{o}}$lder inequality (see Lemma \ref{lem:poincare} and \ref{lem:holder}), we establish the following bounds:
\begin{subequations}\label{eq:bounds_reply}
\begin{align}
    -2D\Vert e^F_x(\cdot, t)\Vert_2 &\leq -2D \Vert  e^F(\cdot, t)\Vert_2^2,
\end{align}
\begin{multline}\label{eq:bound_b}
    -D\left[1-\mathrm{exp}(-K_L t)\right] \int_\mathcal{S} (e^F(x, t))^2 g_1(x) \,\mathrm{d}x \leq\\\leq D\left[1-\mathrm{exp}(-K_L t)\right] \left\vert\int_\mathcal{S} (e^F(x, t))^2 g_1(x) \,\mathrm{d}x \right\vert\leq\\
    \leq D\left[1-\mathrm{exp}(-K_L t)\right] \int_\mathcal{S} \left\vert(e^F(x, t))^2 g_1(x)\right\vert \,\mathrm{d}x =\\= D\left[1-\mathrm{exp}(-K_L t)\right]\Vert (e^F(\cdot, t))^2g_1(\cdot)\Vert_1 \leq\\\leq D\Vert g_1(\cdot)\Vert_\infty \Vert e^F(\cdot, t)\Vert_2^2,
\end{multline}
\begin{multline}\label{eq:bound_c}
    -\mathrm{exp}(-K_L t) \int_\mathcal{S} (e^F(x, t))^2 h_1(x) \,\mathrm{d}x \leq\\
    \leq\mathrm{exp}(-K_L t)\Vert h_1(\cdot)\Vert_\infty  \Vert e^F(\cdot, t)\Vert_2^2,
\end{multline}
\begin{multline}
    2\,\mathrm{exp}(-K_L t)\int_\mathcal{S} e^F(x, t) h_2(x)\,\mathrm{d}x \leq\\\leq 2\,\mathrm{exp}(-K_L t) \left\vert \int_\mathcal{S} e^F(x, t) h_2(x)\,\mathrm{d}x\right\vert \leq\\\leq 2\,\mathrm{exp}(-K_L t) \int_\mathcal{S} \left\vert e^F(x, t) h_2(x)\right\vert \,\mathrm{d}x =\\= 2\,\mathrm{exp}(-K_L t) \Vert e^F(x, t) h_2(x)\Vert_1 \leq\\\leq 2\,\mathrm{exp}(-K_L t) \Vert h_2(x)\Vert_2\Vert e^F(x, t)\Vert_2.
\end{multline}
\end{subequations}
The derivation of \eqref{eq:bound_c} follows the steps of \eqref{eq:bound_b}. Moreover, we took into consideration the fact that $1-\mathrm{exp}(-K_L t)$ is positive and bounded by 1.

Finally, accounting for the bounds in \eqref{eq:bounds_reply} into \eqref{eq:polished_vdot}, yields
\begin{multline}\label{eq:replyVdot_bound}
    \left(\Vert e^F(\cdot, t) \Vert_2^2\right)_t \leq (-2D+D\Vert g_1(\cdot) \Vert_\infty )\Vert e^F(\cdot, t) \Vert_2^2 \\+ \Vert h_1(\cdot)\Vert_\infty \mathrm{exp}(-K_Lt)\Vert e^F(\cdot, t) \Vert_2^2 \\+ 2\Vert h_2(\cdot)\Vert_2 \mathrm{exp}(-K_L t)\Vert e^F(\cdot, t) \Vert_2.
\end{multline}
The bound on the right hand-side of  \eqref{eq:replyVdot_bound} globally converges to 0 due to Lemma \ref{lemma:bounding_sys} (upon setting $\eta = \Vert e^F\Vert_2^2$,  $\beta = -2D+D\Vert g_1 \Vert_\infty$, $\gamma = \Vert h_1\Vert_\infty$, $\delta = 2\Vert h_2\Vert_2$ and $K = K^L$). Hence the comparison Lemma (see Lemma \ref{lemma:comparison_lemma}) yields the claim. 
\end{proof}



The control scheme proposed so far does not rely on any information sensed in real-time about the followers' displacement, rendering the solution not robust to perturbations, as detailed by the numerical simulations reported later in Section \ref{subsec:robustness_analysis}. This underscores the necessity for expanding the strategy in order to incorporate some feedback mechanism on the followers' dynamics, as discussed next.


\section{Reference-governor control}\label{sec:ref_gov}
\begin{figure}
    \centering
    \includegraphics[width=0.5\textwidth]{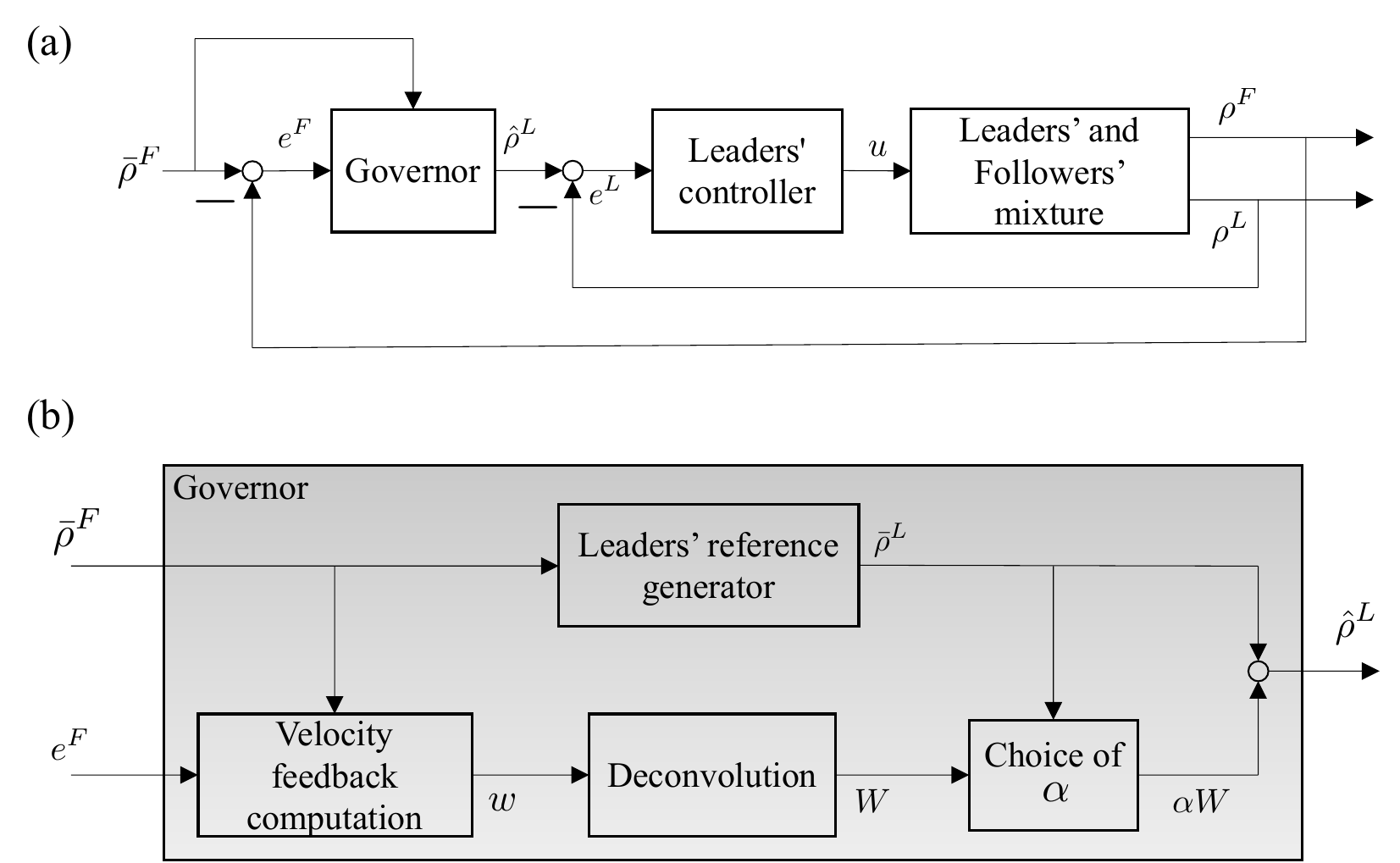}
    \caption{\footnotesize (a) Reference-governor control scheme. (b) Detail of the governor block.}
    \label{fig::control_scheme}
\end{figure}

To incorporate feedback related to the followers' density, we introduce a reference-governor approach, inspired by \cite{bemporad1998reference}. We employ a dual-feedback loop structure: the outer loop, or ``governor loop", dynamically adjusts the target density for the leaders, $\hat{\rho}^L$, aiming to minimize the error $e^F = \bar{\rho}^F-\rho^F$ by facilitating the required organization of the followers. The inner loop, or the leaders' control loop, then calculates the control input $u$ as per \eqref{eq:leaders}, to guide the actual leaders' density, $\rho^L$, towards $\hat{\rho}^L$, thereby reducing the discrepancy $e^L = \hat{\rho}^L - \rho^L$ to zero and fulfilling the control objectives. An illustration of this strategy is depicted in Fig. \ref{fig::control_scheme}.
Note that for controlling the leaders' density we  leverage the framework previously detailed in Section \ref{sec:lead} (see Remark \ref{rem:leaders_tracking} specifically).


\subsection{Governor design} \label{sec:gov_des}
Here, we discuss the design of the governor and explore its stability properties. When leaders are assigned a tracking problem, choosing $u$ in \eqref{eq:leaders} according to \eqref{eq:flux_tv} yields
\begin{align}\label{eq:leaders_sol_track}
    \rho^L(x, t) = \hat{\rho}^L(x, t) + \Psi(x, t),
\end{align}
where $\Psi$ represents the leaders' transient behavior:
    \begin{align}
        \Psi(x, t) = -\left[\hat{\rho}^L(x, 0) + \rho_0^L(x)\right]\mathrm{exp}(-K_L t).
    \end{align}
In establishing \eqref{eq:leaders_sol_track}, we applied Theorem \ref{th:leaders_convergnce} and Remark \ref{rem:leaders_tracking}
Thus, using \eqref{eq:vfl_def} and the linearity of the convolution operator, $v^{FL}$ in \eqref{eq:followers} can be decomposed as
\begin{align}\label{eq:vfl_decomposition}
    v^{FL}(x, t) = \hat{v}^{FL}(x, t) + (f*\Psi)(x, t),
\end{align}
where $\hat{v}^{FL} = f*\hat{\rho}$.
Below, we derive an expression for $\hat{v}^{FL}$ that, through Lyapunov arguments, ensures the convergence of the followers' error to 0.
After that, we show how to deconvolve such a velocity field to obtain $\hat{\rho}^L$, for the leaders to track. Specifically, we choose $\hat{v}^{FL}$ to incorporate the feed-forward action that was discussed in Section \ref{sec:control design}, and a feedback correction whose weight can be chosen online to ensure physical constraints are met. In particular,
    \begin{align}\label{eq:vfl_ff_fb}
        {\hat{v}^{FL} (x, t) =  (f*\hat{\rho}^L)(x, t)} = \bar{v}^{FL}(x) + \alpha(t) \,w(x, t) 
    \end{align}
where $\bar{v}^{FL}$ is the feed-forward term chosen as in  \eqref{eq:vfl_bar}, $\alpha: \mathbb{R}_{\geq 0} \rightarrow [0,1]$ is a control function to be appropriately selected,  and
\begin{align}\label{eq:w}
    w(x, t) =  \frac{D\bar{\rho}^F_x(x)e^F(x, t)}{\bar{\rho}^F(x)\left(\bar{\rho}^F(x)- e^F(x,t)\right)}.
\end{align}
is a feedback correction term modulated by $\alpha(t)$. 

\begin{rem}
Equation \eqref{eq:w} is well defined only if $\rho^F > 0$, as $\bar{\rho}^F\in\mathbb{R}_{>0}$ (see Section \ref{sec:prob_statement}). This condition is reasonable as: (i) there is no need to exert a control action where $\rho^F = 0$, and (ii) in practical scenarios, $\rho^F$ is estimated from the positions of a discrete set of agents using, for instance, a Gaussian kernel estimator \cite{silverman2018density}, ensuring this assumption is met. 
\end{rem}

\begin{theorem} [Followers' global stability]\label{th:foll_conv}
    Let the followers' density evolve according to \eqref{eq:followers}, with $v^{FL}$ given by \eqref{eq:vfl_decomposition} and $\hat{v}^{FL}$ by \eqref{eq:vfl_ff_fb}. If the feasibility condition in Theorem \ref{proposition:prop1} holds and $\Vert g_1(\cdot)\Vert_\infty < 2$ (see \eqref{eq:g1}), then, for any choice of $\alpha(t) \in [0,1]$, the followers' density converges globally to the desired density $\bar{\rho}^F$ in $\mathcal{L}^2(\mathcal{S})$.
\end{theorem}
\begin{proof}
    The followers' error dynamics obeys  \eqref{eq:err_dyn_act_stig} with initial and periodic boundary conditions set as in \eqref{eq:IC-PBC_err}. Substituting \eqref{eq:vfl_decomposition} into \eqref{eq:err_dyn_act_stig} yields
    \begin{multline}
    e^F_t(x, t) = D \left(e_{xx}^F(x, t)-\bar{\rho}^F_{xx}(x)\right)\\
    \begin{aligned}
        &+\left[\left(\bar{\rho}^F(x)-e^F(x, t)\right) \hat{v}^{FL}(x,t)\right]_x \\    &+\left[\left(\bar{\rho}^F(x)-e^F(x, t)\right) (f*\Phi)(x, t)\right]_x.
    \end{aligned}
    \end{multline}
    Substituting \eqref{eq:vfl_ff_fb} (with $w$ from \eqref{eq:w}), we obtain 
    \begin{multline} 
        \begin{aligned}
        &e^F_t(x, t) = De^F_{xx}(x, t) \\
            &- D \left[1-\mathrm{exp}(-K_L t)\right]\left[e^F(x, t)\frac{\bar{\rho}^F_x(x)}{\bar{\rho}^F(x)}\right]_x\\&+ D\alpha(t)\left[e^F(x, t)\frac{\bar{\rho}^F_x(x)}{\bar{\rho}^F(x)}\right]_x+\mathrm{exp}(-K_L t)\times
        \end{aligned}\\
        \times\left[(\bar{\rho}^F(x)-e^F(x, t)) (f*\rho^L_0)(x) -D\bar{\rho}^F_{x}(x)\right]_x.
    \end{multline}
    Similarly to Theorem \ref{th_act_stig_stab}, we introduce the Lyapunov functional $\Vert e^F\Vert_2^2$; differentiating in time, yields
    \begin{multline}
        \left(\Vert e^F(\cdot, t) \Vert_2^2\right)_t = 2 D \int_\mathcal{S} e^F(x, t) e^F_{xx}(x, t) \,\mathrm{d}x \\- 2 D \left[1-\mathrm{exp}(-K_L t)\right] \int_\mathcal{S} e^F(x, t)\left[e^F(x, t) \frac{\bar{\rho}^F_x(x, t)}{\bar{\rho}^F(x)}\right]_x  \,\mathrm{d}x \\+2D\alpha(t)\int_\mathcal{S} e^F(x, t)\left[e^F(x, t) \frac{\bar{\rho}^F_x(x, t)}{\bar{\rho}^F(x)}\right]_x  \,\mathrm{d}x 
    +2\mathrm{exp}(-K_L t)\times \\\times\int_{\mathcal{S}} e^F(x, t) \left[(\bar{\rho}^F(x)-e^F(x, t)) (f*\rho^L_0)(x)-D\bar{\rho}^F_{x}(x)\right]_x\mathrm{d}x.
    \end{multline}
    By means of integration by parts and recalling that $[(e^F)^2]_x = 2e^Fe^F_x$, we establish the following identity:
    \begin{multline}
         2 D \alpha(t)\int_\mathcal{S} e^F(x, t)\left[e^F(x, t) \frac{\bar{\rho}^F_x(x, t)}{\bar{\rho}^F(x)}\right]_x  \,\mathrm{d}x =\\= -D\alpha(t) \int_\mathcal{S} 2e^F(x, t)e^F_x(x, t) \frac{\bar{\rho}^F_x(x, t)}{\bar{\rho}^F(x)} \,\mathrm{d}x =\\= -D\alpha(t) \int_\mathcal{S} \left[(e^F(x, t))^2\right]_x \frac{\bar{\rho}^F_x(x, t)}{\bar{\rho}^F(x)} \,\mathrm{d}x=\\=
    D\alpha(t) \int_\mathcal{S} (e^F(x, t))^2 g_1(x) \,\mathrm{d}x.
    \end{multline}
    Using this identity along with those in \eqref{eq:identities}, we obtain
    \begin{multline}
        \left(\Vert e^F(\cdot, t) \Vert_2^2\right)_t = -2D\Vert e^F_x(\cdot, t) \Vert_2^2 \\-D\left[1-\mathrm{exp}(-K_L t)-\alpha(t)\right] \int_\mathcal{S} (e^F(x, t))^2 g_1(x) \,\mathrm{d}x \\-\mathrm{exp}(-K_L t)\int_\mathcal{S}(e^F(x, t))^2 h_1(x)\,\mathrm{d}x \\+ 2\mathrm{exp}(-K_L t)\int_\mathcal{S} e^F(x, t) h_2(x)\,\mathrm{d}x.
    \end{multline}
    As in \eqref{eq:bound_b}, we establish the following bound:
    \begin{figure*}
     \centering
     \begin{subfigure}[b]{0.32\textwidth}
         \centering
         \includegraphics[width=\textwidth]{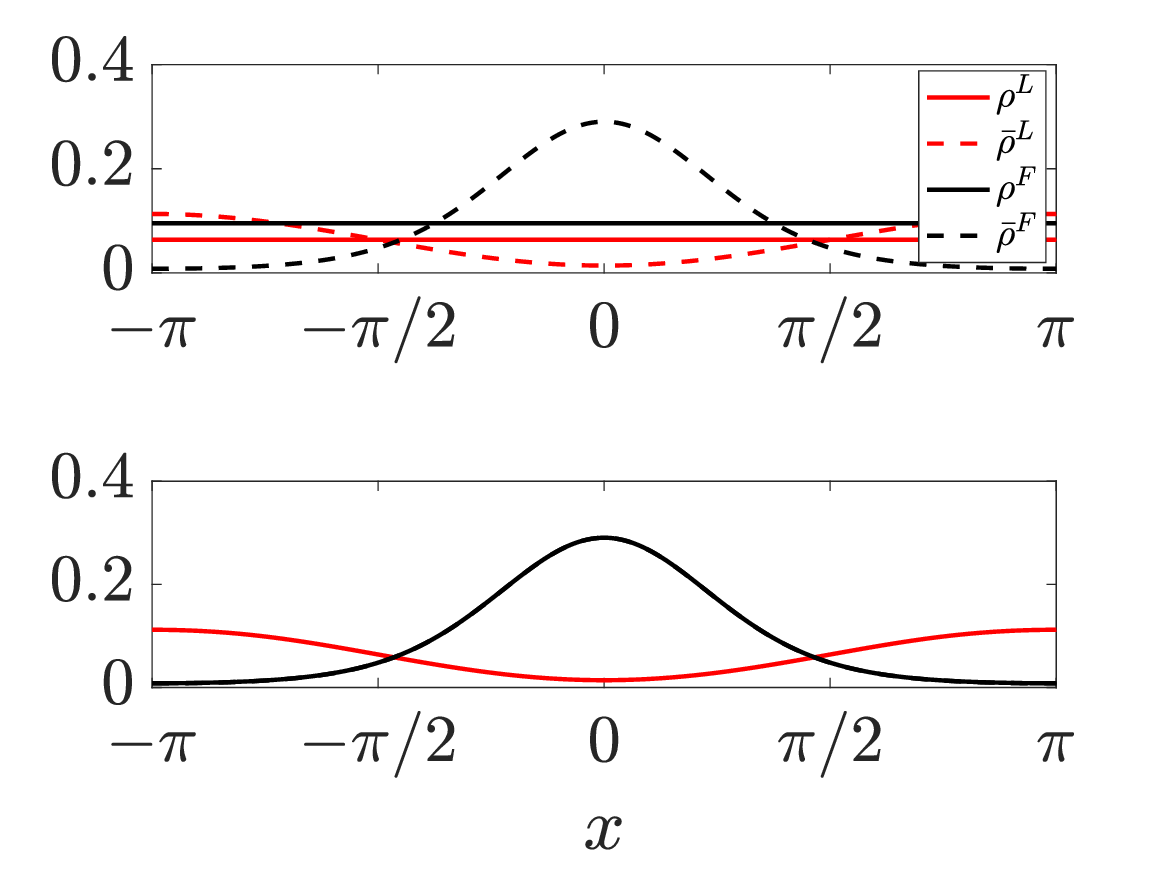}
         \caption{}
         \label{sub11}
     \end{subfigure}
     \begin{subfigure}[b]{0.32\textwidth}
         \centering
         \includegraphics[width=\textwidth]{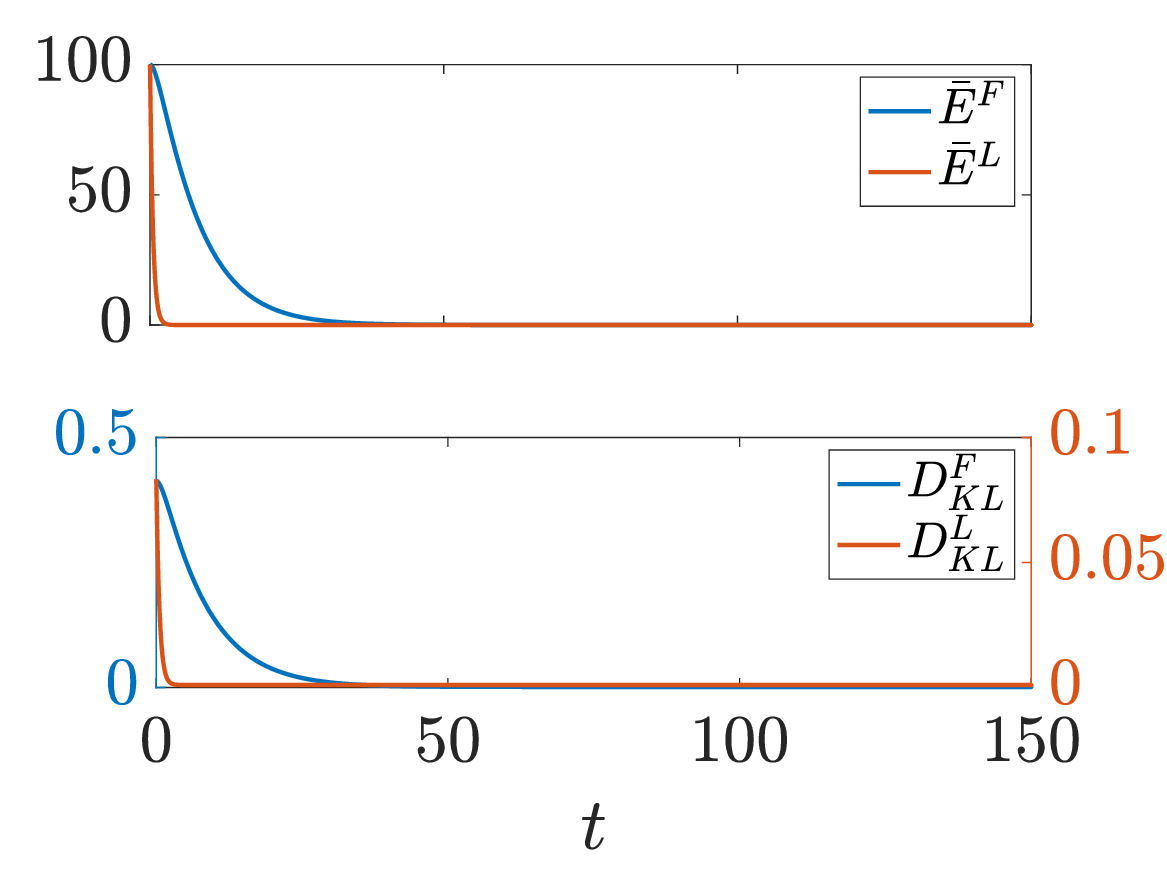}
         \caption{}
         \label{sub12}
     \end{subfigure}
     \begin{subfigure}[b]{0.32\textwidth}
         \centering
         \includegraphics[width=\textwidth]{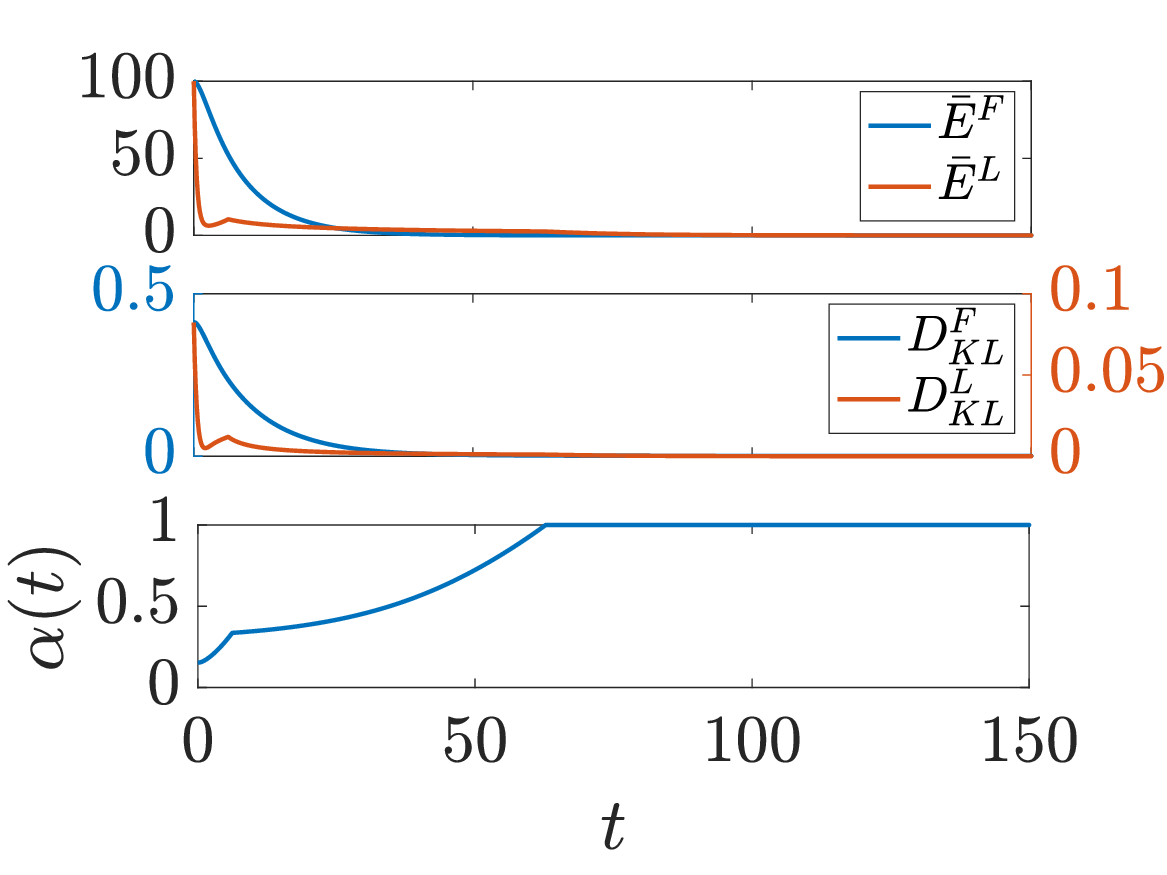}
         \caption{}
         \label{sub13}
     \end{subfigure}
        \caption{\footnotesize Monomodal trial: (a) initial and final densities; (b) time evolution of the percentage error and KL divergences using the feed-forward control scheme; and (c) time evolution of the percentage error, KL divergences, and $\alpha$ using the reference-governor scheme.}
        \label{fig:centered_vonMises}
\end{figure*}
\begin{multline}\label{eq:bound_modified}
         -D\left[1-\mathrm{exp}(-K_L t)-\alpha(t)\right] \int_\mathcal{S} (e^F(x, t))^2 g_1(x) \,\mathrm{d}x \leq\\
         \leq D\Vert g_1(\cdot)\Vert_\infty \Vert e^F(\cdot, t)\Vert_2^2,
    \end{multline}
    where we used H$\ddot{\mathrm{o}}$lder inequality and the fact that, for any $\alpha(t) \in [0, 1]$, 
        $\left\vert1-\mathrm{exp}(-K_L t)-\alpha(t)\right\vert <1, \;\forall t \in \mathbb{R}_{\geq 0}$.
    Combining bound \eqref{eq:bound_modified} with the bounds in \eqref{eq:bounds_reply}, we establish that 
    \begin{multline}
         \left(\Vert e^F(\cdot, t) \Vert_2^2\right)_t \leq (-2D+D\Vert g_1(\cdot) \Vert_\infty )\Vert e^F(\cdot, t) \Vert_2^2 \\+ \Vert h_1(\cdot)\Vert_\infty \mathrm{exp}(-K_Lt)\Vert e^F(\cdot, t) \Vert_2^2 \\+ 2\Vert h_2(\cdot)\Vert_2 \mathrm{exp}(-K_L t)\Vert e^F(\cdot, t) \Vert_2.
    \end{multline}
    Then, as in Theorem \ref{th_act_stig_stab}, using Lemma \ref{lemma:bounding_sys} proves the claim.
\end{proof}
\begin{rem}
    The case $\alpha = 0$ coincides with the control technique studied in Section \ref{sec:control design}. 
\end{rem}

Given \eqref{eq:vfl_decomposition} and \eqref{eq:vfl_ff_fb}, and recalling that $\hat{v}^{FL} = f*\hat{\rho}^L$, we recover the desired leaders' density $\hat{\rho}^{L}$ by online deconvolution \cite{wing1991primer} of $v^{FL}$ with the repulsive interaction kernel given by \eqref{eq:rep_kern} (see Appendix \ref{app:deconv} for more details). For the linearity of the convolution, we can deconvolve the two terms of \eqref{eq:vfl_ff_fb} separately, leading to
\begin{align}\label{eq:rhohat_deconv}
    \hat{\rho}^L(x, t) = \bar{\rho}^L(x) + \alpha(t) W(x, t),
\end{align}
where $\bar{\rho}^L$ is the deconvolution of $\bar{v}^{FL}$ and $W$ is the deconvolution of $w$, that is,
\begin{align}\label{eq:Wdef}
    W(x, t) = \frac{w_x(x, t)}{2}-\frac{1}{2L^2}\int w(x, t)\,\mathrm{d}x + \beta(t),
\end{align}
with $\beta$ being an arbitrary function of time (see Appendix \ref{app:deconv} for more details).
Being the problem feasible, we know that $\bar{\rho}^L$ is positive and sums to $M^L$. Then, for $\hat{\rho}^L$ to be physically meaningful, that is, positive and summing to $M^L$, $W$ needs to fulfill the following conditions:
\begin{subequations}
\begin{align}
    \int_\mathcal{S} W(x, t) \,\mathrm{d}x = 0&, \;\;\forall t\geq 0,\label{eq:Wpositity}\\
    \bar{\rho}^L(x) + \alpha(t) W(x, t) \geq 0&, \;\; \forall t\geq 0, x \in \mathcal{S}\label{eq:Wsum}.
\end{align}
\end{subequations}
Condition \eqref{eq:Wpositity} can always be ensured by appropriately choosing $\beta$ in \eqref{eq:Wdef} and \eqref{eq:Wsum} can be satisfied by selecting  $\alpha$ so  that it remains fulfilled, as will be shown next. 

\subsubsection*{Choice of $\alpha(t)$}\label{sec:alpha_choice}
A possible conservative choice is to set
\begin{align}\label{eq:alpha_choice}
    \alpha(t) = \left[ \frac{-\min_x \bar{\rho}^L(x)}{\min_x W(x, t)} \right]_0^1,
\end{align}
where subscripts and superscripts of square brackets indicate a saturation.
With this choice of $\alpha(t)$, we can guarantee that
\begin{align}\label{eq:worst_case_positivity}
    \min_x \bar{\rho}^L(x) + \alpha(t) \min_x W(x, t) \geq 0,
\end{align}
and therefore that \eqref{eq:Wsum} is fulfilled.
Note that in making the choice we exploited the fact that $\min_x W \leq 0$ by construction, since $\beta$  is chosen in \eqref{eq:Wdef} to ensure \eqref{eq:Wpositity}. Also, notice that  \eqref{eq:worst_case_positivity} and \eqref{eq:Wsum} remain satisfied when $\alpha$ is saturated to zero as $\bar \rho^L \geq 0$ by assumption, and when $\alpha$ is saturated to unity as $-\min_x \bar{\rho}^L / \min_x W > 1$ implies $\vert \min_x \bar{\rho}^L \vert > \vert \min_x W\vert$.

Other possible choices of $\alpha$, including optimal ones, are possible.
A practical heuristic choice to approximate the optimal $\alpha$ and enhance robustness of the algorithm to persistent disturbances is adopted in Section \ref{subsec:robustness_analysis}. 


\section{Numerical validation}\label{sec:num_val1}
In this section, we perform a numerical validation of the two proposed control strategies. For the numerical integration of \eqref{eq:leaders}-\eqref{eq:followers}, we use a central finite difference scheme with a mesh of 500 cells, and we approximate time derivatives with Forward Euler with a fixed time step $\mathrm{d}t = 0.001$.

For each trial, we consider $D=0.05$, $L=\pi$, and a time horizon of 150,000 time iterations and recorded followers and leaders percentage error, that is,
\begin{align}
    \bar{E}^i(t) = \frac{\Vert e^i(\cdot, t)\Vert_2^2}{\max_t \Vert e^i(\cdot, t)\Vert_2^2}\,100,\;\;i=F, L.
\end{align}
As an extra performance index, we borrow from the optimal transport literature \cite{chen2021optimal} the Kullback-Leibler (KL) divergence \cite{kullback1951information} (or relative entropy) between the desired followers' (leaders') density and the followers' (leaders) density, that is
\begin{align}
    D^i_{KL}(t) = \int_{\mathcal{S}} \bar{\rho}^i(x) \log \left(\frac{\bar{\rho}^i(x)}{\rho^i(x, t)}\right)\mathrm{d}x,\;\;i=F, L.
\end{align}

We study a monomodal regulation problem. Specifically, we set $M^L = 0.4$ and choose the von Mises distribution in \eqref{eq:vonmises} with $\kappa = 1.8$ and $\mu = 0$ for the desired followers' density. We report the results of the numerical example in Fig. \ref{fig:centered_vonMises}. Specifically, we show the initial and final displacement of the leaders' and followers' densities, resulting in the same steady-state profile with both the control techniques. Then, in Fig. \ref{sub12} and \ref{sub13} (upper panel), we report the time evolution of the percentage errors and KL divergences using respectively the feed-forward and the reference-governor control schemes. In Fig. \ref{sub13} (bottom panel), we show the time evolution of the control function $\alpha$ selected according to \eqref{eq:alpha_choice}.
Similar results were obtained for bi-modal regulation tasks (see the Supplementary material available at \cite{supp_material}).

\subsection{Robustness analysis}\label{subsec:robustness_analysis}
\begin{figure}
     \centering
         \includegraphics[width=0.3\textwidth]{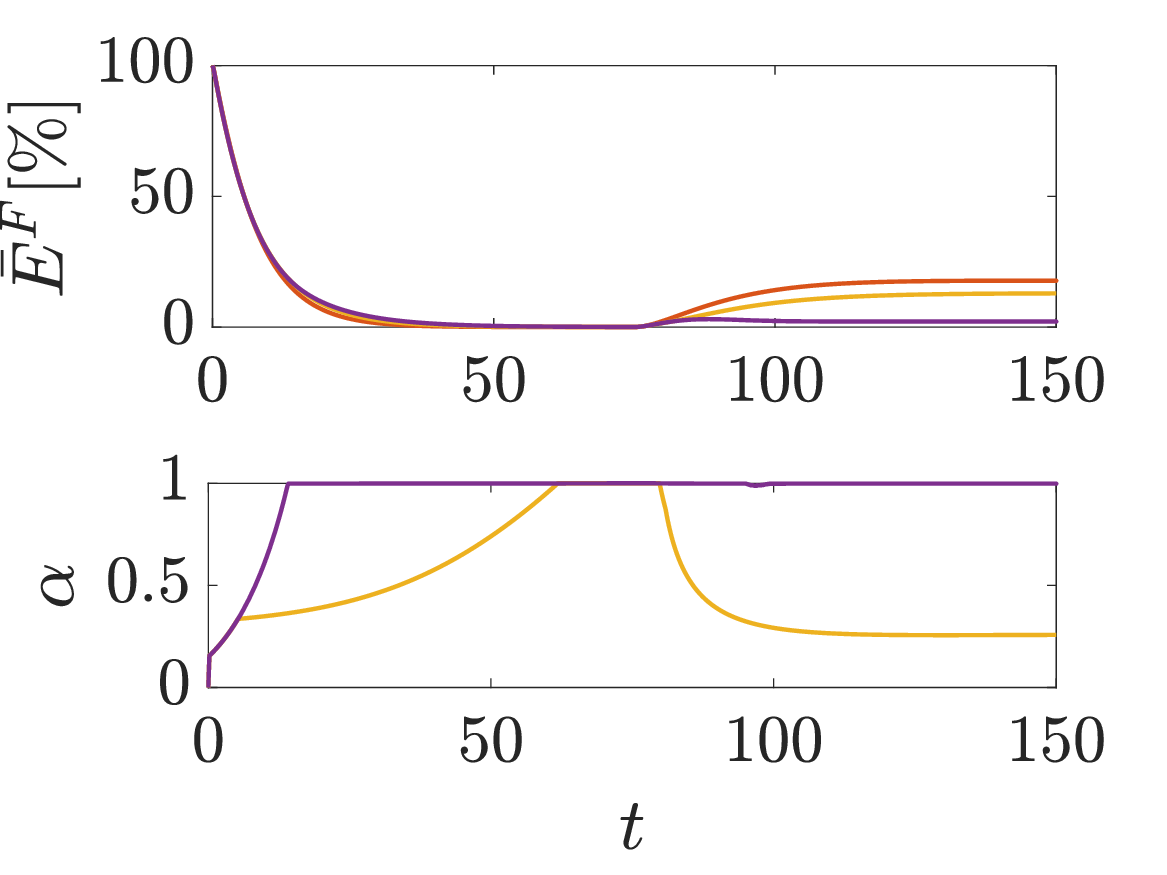}
         
        \caption{ \footnotesize Robustness to external disturbance. Percentage error (top panel) and evolution of $\alpha$ (bottom panel) in time for the feedback control schemes (orange line for feed forwards, yellow line for reference-governor and purple line for reference-governor with an improved choice of $\alpha$).}
        \label{subfig:err_alpha}
\end{figure}
To underscore the benefits of the strategy incorporating the reference governor control over the simpler feed-forward control strategy, we proceed to examine the robustness of both strategies against disturbances and structural perturbations. Our findings demonstrate that, as anticipated, the strategy equipped with the reference governor control offers superior compensation for these disruptions. 

\subsubsection{Perturbations}
To begin, we consider the dynamics of the followers to be perturbed by an additive velocity field $d$, defined as
\begin{align}
d(x, t) = \frac{\pi}{100} \, \mathrm{step}(t -t_\mathrm{f}/2).
\end{align}
This represents a positive constant drift which is suddenly introduced into the followers' dynamics halfway through a simulation trial.

Considering the same setting as the one depicted in Fig. \ref{fig:centered_vonMises} where the goal is for the followers to achieve a monomodal distribution, we applied both the feed-forward and the reference-governor schemes, observing enhanced performance with the latter (see Fig. \ref{subfig:err_alpha}). 
Specifically, as illustrated in the top panel of Fig. \ref{subfig:err_alpha}, the steady-state percentage residual error decreases from nearly 20\% to approximately 10\% with the introduction of feedback. 

Performance improves more significantly, when we introduce a numerical procedure to improve the choice of $\alpha$. 
Specifically, the optimal $\alpha$ (that is, the maximum value still fulfilling \eqref{eq:Wsum}) can be formalized as
\begin{align}\label{eq:opt_alpha}
    \alpha(t) = \lim_{\varepsilon\to0+}\left[\min_x \left(\frac{\bar\rho^L(x)}{\max\{-W(x,t), \varepsilon\}} \right)\right]_0^1,
\end{align}
which we practically implement by fixing $\varepsilon = 0.01$.
Specifically, as $\alpha$ remains set to unity for extended periods (as shown in the bottom panel of Fig. \ref{subfig:err_alpha}), the feedback correction intensifies, leading to a residual percentage error of only 2\% (as detailed in the top panel of Fig. \ref{subfig:err_alpha}).

\begin{figure}
    \centering
    \includegraphics[width=0.3\textwidth]{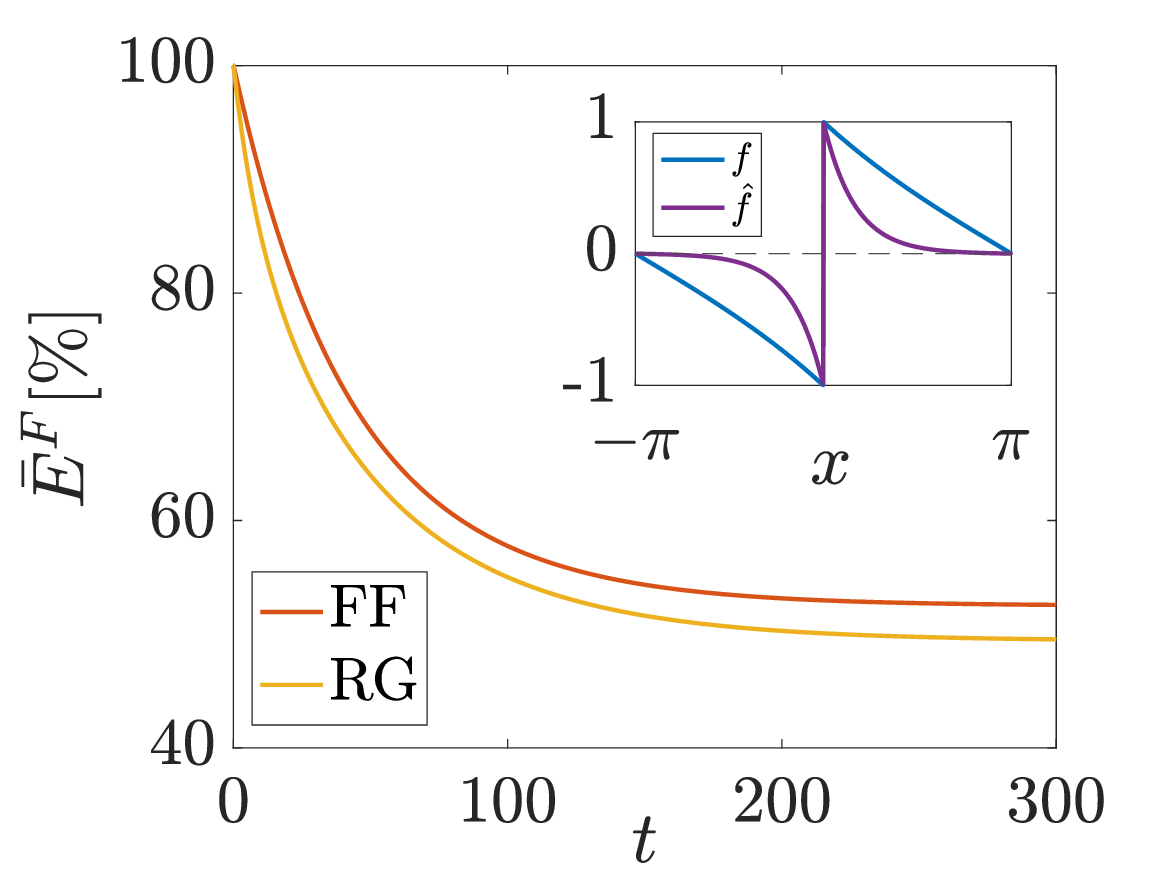}
    \caption{\footnotesize Robustness to uncertainties: time evolution of the percentage error with the feed-forward (FF) control scheme and the reference-governor (RG) control scheme. In the inset, the nominal and perturbed interaction kernel.}
    \label{fig::uncertainties}
\end{figure}

\subsubsection{Robustness to Structural Perturbations}
To evaluate robustness to structural perturbations, we assess the response to parametric uncertainties in the characteristic length scale $L$ of the interaction kernel $f$ (see \eqref{eq:rep_kern}). This involves assuming a discrepancy between the nominal length scale used for control design and the actual scale influencing the followers' dynamics. Specifically, setting $D=0.02$ and using the same monomodal configuration depicted in Fig. \ref{fig:centered_vonMises}, we compute $u$ in \eqref{eq:leaders} using the nominal value $L = \pi$ for both the feed-forward and reference-governor schemes. Conversely, in the numerical simulations, the followers are assumed to react to the leaders' displacement through a perturbed kernel $\hat{f}$, defined as \eqref{eq:rep_kern} with $L = \pi/6$. The results of this trial, along with graphical representations of both the nominal and perturbed kernels, are presented in Fig. \ref{fig::uncertainties}. We find that that the reference-governor scheme enhances steady-state performance, reducing the steady-state percentage error to almost 45\% as compared to 55\% observed when the feed-forward scheme is adopted. We did not document the time evolution of $\alpha$ because, adhering to the conservative approach outlined in Section \ref{sec:alpha_choice}, we fixed it at 1 throughout the trial.

In the presence of parametric uncertainties on the diffusion coefficient $D$ (omitted here for brevity), the behavior of both the feed-forward and reference-governor schemes remains similar. This similarity arises because the feedback action $w$ is not independent of $D$, as illustrated in \eqref{eq:w}.

\section{An application to multi-agent leader-follower systems via continuification}\label{sec:herding}
\begin{figure*}
     \centering
     \begin{subfigure}[b]{0.32\textwidth}
         \centering
         \includegraphics[width=\textwidth]{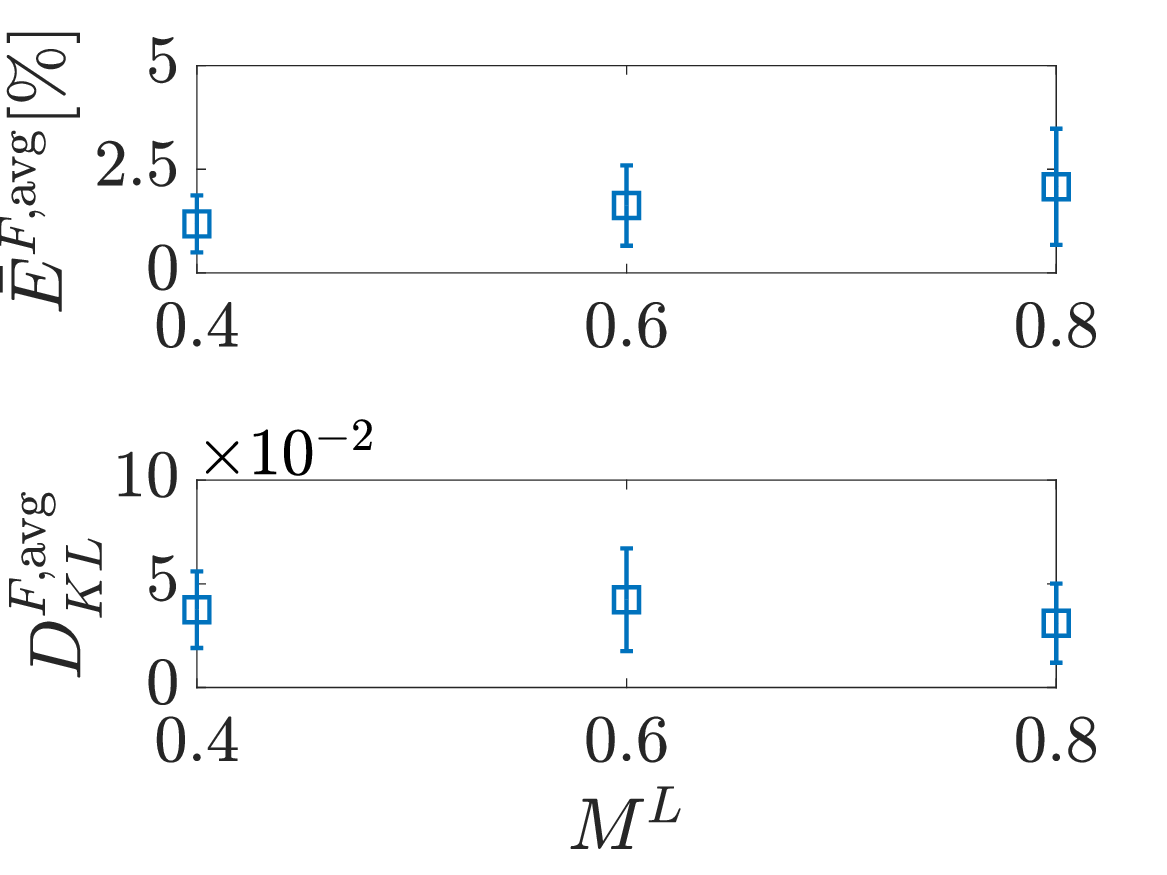}
         \caption{}
         \label{subfig:aggregated}
     \end{subfigure}
     \begin{subfigure}[b]{0.32\textwidth}
         \centering
         \includegraphics[width=\textwidth]{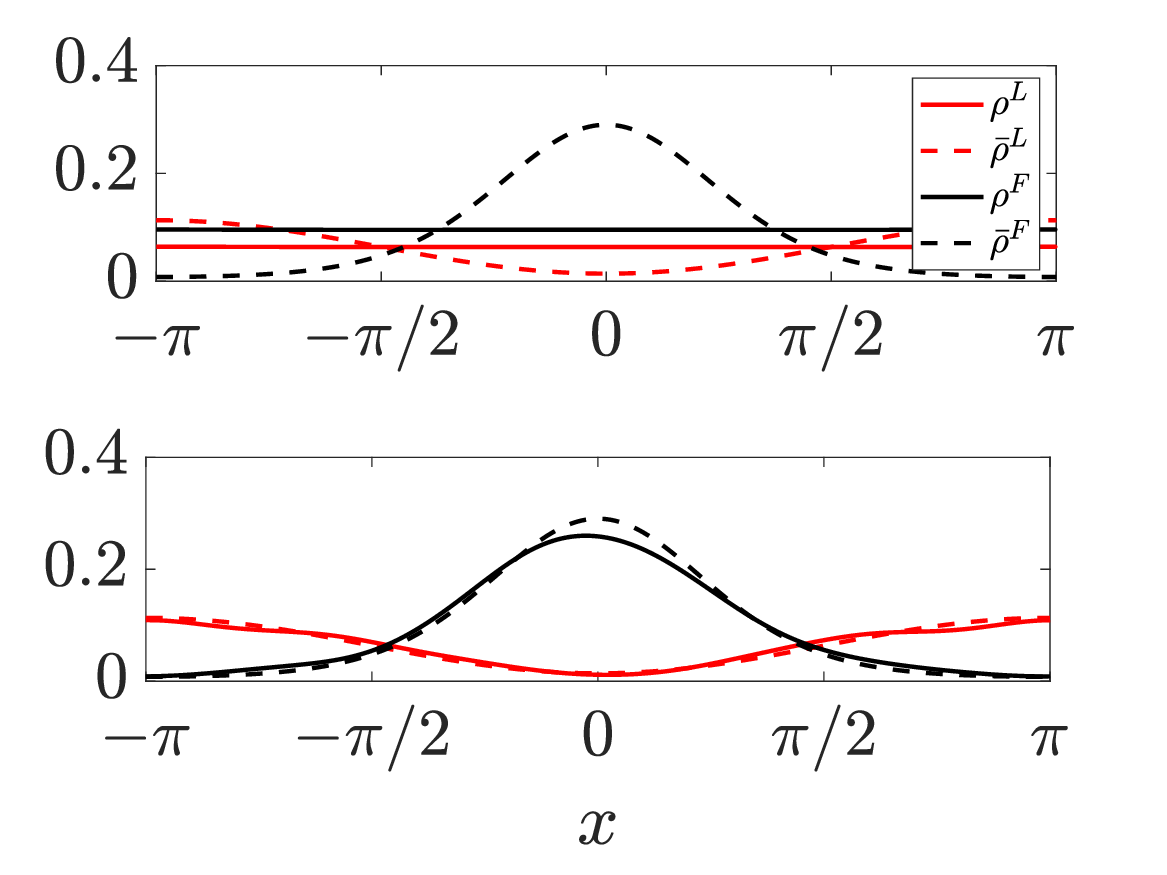}
         \caption{}
         \label{subfig:config_herding}
     \end{subfigure}
     \begin{subfigure}[b]{0.32\textwidth}
         \centering
         \includegraphics[width=\textwidth]{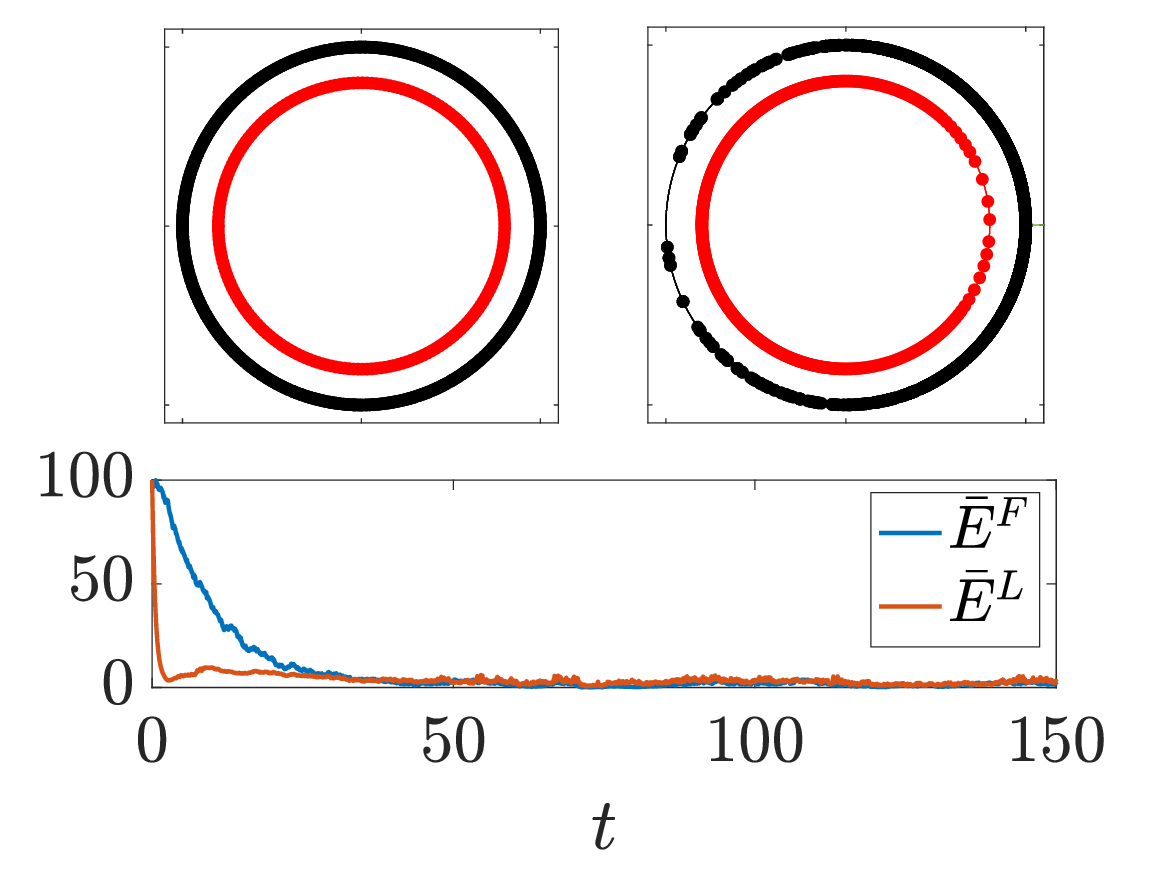}
         \caption{}
         \label{subfig:KL_divs}
     \end{subfigure}
        \caption{\footnotesize Discrete trial: (a) average followers' percentage ($\pm$ one standard deviation) error and KL divergence for different values of the leaders' mass; (b) initial and final densities for a single trial ($N^L= 400$); (c)-(upper panel) initial and final agents displacements in $\mathcal{S}$ for a single trial ($N^L= 400$); and (c)-(lower panel) percentage error of leaders and followers in time for a single trial ($N^L= 400$).}
        \label{fig:herding}
\end{figure*}
Within the framework of continuification-based control approaches \cite{maffettone2022continuification, maffettone2023hybrid, nikitin2021continuation}, the goal is to design microscopic control inputs to influence the  macroscopic dynamics of multi-agent systems, using a continuum approximation.

To validate the macroscopic control solution proposed in this work, we consider a discrete set of stochastic differential equations that replicate the leader-follower scenario previously examined. In particular, we assume a population of $N^L$ leaders needs to steer the dynamics of a population of $N^F$ followers. We consider the two populations move in $\mathcal{S}$ and, as often assumed in the literature, e.g. \cite{lama2023shepherding}, we set their dynamics as
\begin{subequations}\label{eq:discrete_model}
    \begin{align}
        \dot{x}^L_i &= u_i, \;\;i=1, \dots, N^L \label{eq:disc_lead}\\
        \mathrm{d}{x}_k^F &= \frac{1}{N^L+N^F}\sum_{j=1}^{N^L} f({x_j^L\triangleright x_k^F})\,{\mathrm{d}t} + \sqrt{2D}\mathrm{d}B_k,  \nonumber\\&\qquad\qquad\qquad\qquad\qquad\qquad k=1, \dots, N^F,
\end{align}
\end{subequations}
where $B_k$ is a standard Wiener process. Such a formulation represents the discrete counterpart of \eqref{eq:leaders}-\eqref{eq:followers} \cite{almi2023}. Following our solution, and in the context of a continuification scheme, we can estimate the density of the group starting from the agents' positions and perform a discretization. {This consists in fixing } the microscopic control inputs of the leaders $u_i$ in \eqref{eq:disc_lead} as
\begin{align}
u_i(t) = u(x_i, t), \;\;i=1, \dots, N^L,
\end{align}
with $u$ coming from \eqref{eq:flux_tv}, and considering the reference governor scheme proposed in Section \ref{sec:ref_gov}. 
Our proposed discretization procedure consists of spatially sampling a macroscopic control field, differing from \cite{nikitin2021continuation}, as we do not require agents to interact on a spatially invariant network topology.

We consider the discrete counterpart of the numerical setup discussed in Section \ref{sec:num_val1}. Specifically, we set $D = 0.05$, $L = \pi$, $K_L = 1$, and simulate a total of $N^L + N^F = 1000$ agents. For the desired followers' density, we adopt the monomodal von Mises distribution utilized in the trial depicted in Fig. \ref{fig:centered_vonMises}. Agent densities are estimated from their positions using an ad-hoc Gaussian kernel estimation method \cite{silverman2018density}, and numerical integration is performed using the forward Euler method for leaders and the Euler-Maruyama method for followers, with a time step of $\Delta t = 0.001$.

We fix the initial densities of both populations to be constant and conduct $n=128$ trials, each consisting of 150,000 time steps, while exploring different feasible ratios of leaders to followers. We characterize this numerical investigation by calculating the average over the $n$ trials of the steady-state percentage error and KL divergence. Specifically, results are depicted in Fig. \ref{subfig:aggregated}, where it is evident that we consistently reduce the percentage error to well below 5\% -- a performance level also corroborated by the KL divergence of the followers. For completeness, Fig. \ref{subfig:config_herding} and \ref{subfig:KL_divs} also present the outcomes of a single trial with $N^L = 400$, in terms of densities, agents' displacements, and percentage error. 

Contrary to the macroscopic simulations performed using the continuum formulation in Section \ref{sec:num_val1}, we register a small steady-state error in the discrete model. This error primarily arises from two factors: the finite size of the swarm, which, practically, constrains the continuum hypothesis to hold only partially, and the stochastic behavior of the followers.

\section{Extension to higher dimensions}\label{sec:extension}
Our one-dimensional framework can be readily extended to higher dimensions. Specifically, assuming the spatial domain to be $\Omega:=[-\pi, \pi]^d$ (with $d=2, 3$), in $H^2(\Omega)$, the model becomes
\begin{subequations}
    \begin{align}
    \rho^L_t(\mathbf{x}, t) + \nabla \cdot \left[\rho^L(\mathbf{x}, t) \mathbf{u}(\mathbf{x}, t)\right] &= 0\label{eq:leaders_Rd},\\
    \rho^F_t(\mathbf{x}, t) + \nabla \cdot \left[\rho^F(\mathbf{x}, t) \mathbf{v}^{FL}(\mathbf{x}, t)\right] &= D\nabla^2\rho^F(\mathbf{x}, t)\label{eq:followes_Rd},
\end{align}
\end{subequations}
where  $\nabla \cdot (\cdot)$ and $\nabla^2(\cdot)$ are the divergence and Laplacian operators, respectively, $\mathbf{x}\in\Omega$, $\mathbf{u}$ is the control input to be designed, and
\begin{align}\label{eq:vfl_Rd_def}
    \mathbf{v}^{FL}(\mathbf{x}, t) = \int_\Omega \mathbf{f}(\mathbf{y}\triangleright\mathbf{x}) \rho^L(\mathbf{y}, t) \,\mathrm{d}\mathbf{y} = (\mathbf{f}*\rho^L)(\mathbf{x}, t),
\end{align}
is the circular convolution of $\rho^L$ with $\mathbf{f}$, that is the $d$-dimensional repulsive interaction kernel.
\begin{rem}
    Notice that a closed form expression for $\mathbf{f}$ was not found. The periodized kernel can be expressed as an infinite series (see \eqref{eq:q_T_def} for the one-dimensional counterpart), which can be truncated for implementation purposes.
\end{rem}
Similarly to the one-dimensional case, to ensure mass is conserved, $\rho^F$ and $\rho^L$ are assumed to be periodic on $\partial\Omega$,  and initial conditions are set similarly to \eqref{eq:leaders_init} and \eqref{eq:foll_init}. Moreover, the total masses of leaders and followers are such that $M^F+M^L = 1$. 

\subsection{Feasibility analysis}\label{sec:feasib_high_d}
Given the problem statement in Section \ref{sec:prob_statement}, we seek the desired velocity field for the followers by assuming that $\bar{\rho}^F$ solves \eqref{eq:followes_Rd} at steady-state
\begin{align}
\nabla \cdot \left[\bar{\rho}^F(\mathbf{x}) \bar{\mathbf{v}}^{FL} (\mathbf{x})\right] = D\nabla^2 \bar{\rho}^F(\mathbf{x}).
\end{align}
Unlike the one-dimensional case, this scalar relation alone does not suffice to uniquely determine the vector field $\bar{\mathbf{v}}^{FL}$. Thus, following the approach in \cite{maffettone2023hybrid}, we define $\mathbf{w} = \bar{\rho}^F \bar{\mathbf{v}}^{FL}$ and impose an irrotationality condition, leading to
\begin{align}\label{eq:electro_static_field}
    \begin{cases}
        \nabla\cdot\mathbf{w}(\mathbf{x}) = D\nabla^2\bar{\rho}^F(\mathbf{x}),\\
        \nabla \times \mathbf{w}(\mathbf{x}) = 0,
    \end{cases}
\end{align}
with periodic boundary conditions applied to $\mathbf{w}$ (in two dimensions the curl operator returns a three dimensional vector with null third component -- for more details see \cite{griffiths2023introduction}, Sec. 1.2.5, example 1.5). 
Being $\mathbf{w}$ irrotational and $\Omega$ simply connected, we conclude that $\mathbf{w} = - \nabla\varphi$, where $\varphi$ is an unknown scalar potential. 

Using this expression of $\mathbf{w}$, equation \eqref{eq:electro_static_field} simplifies into the Poisson equation
\begin{align}
    \nabla^2\varphi(\mathbf{x}) = - D\nabla^2\bar{\rho}^F(\mathbf{x})
\end{align}
which is fulfilled choosing $\varphi = -D \bar{\rho}^F$. With this definition of $\mathbf{w}$ and $\varphi$, we obtain
\begin{align}\label{eq:vfl_bar_Rd}
    \bar{\mathbf{v}}^{FL} (\mathbf{x}) = D\frac{\nabla\bar{\rho}^F(\mathbf{x})}{\bar{\rho}^F(\mathbf{x})}, 
\end{align}
which is the $d$-dimensional extension of \eqref{eq:vfl_bar}. With the additional irrotationality constraint of the flux $\mathbf{w}$, the higher-dimensional formulation is analogous to the one-dimensional one.

Recalling that $\bar{\mathbf{v}}^{FL} = \mathbf{f} * \bar{\rho}^L$, we derive $\bar{\rho}^L$ by deconvolution, 
\begin{align}\label{eq:deconv_Rd}
\bar{\rho}^L(\mathbf{x}) = H(\mathbf{x}) + A,
\end{align}
where $A$ is an arbitrary constant. This deconvolution of $\bar{\rho}^L$ is defined up to an arbitrary constant due to the linearity of the convolution operator and the assumption that the kernel is odd and periodic. Unlike the one-dimensional case, where $\mathbf{f}$ has a closed form, $H$ can only be computed numerically \cite{wing1991primer}. Consequently, the feasibility problem is reformulated in terms of the constant $A$.
\begin{prop}\label{prop:feasib2D}
The problem outlined by \eqref{eq:leaders_Rd}, \eqref{eq:followes_Rd}, and \eqref{eq:problem_statement} admits a feasible solution if there exists a value of $A$ in \eqref{eq:deconv_Rd} such that
\begin{subequations}
\begin{align}
\bar{\rho}^L(\mathbf{x}) &\geq 0, \label{eq:pos_const_2D}\\
\int_\Omega \bar{\rho}^L(\mathbf{x})\,\mathrm{d}\mathbf{x} &= M^L\label{eq:mass_const_2D}.
\end{align}
\end{subequations}
\end{prop}

Proposition \ref{prop:feasib2D} can be evaluated numerically in straightforward steps. One can set $A = a_1 + a_2$ in \eqref{eq:deconv_Rd}, with $a_1$ chosen to minimize its integral, that is, $a_1 = -\min_\mathbf{x} H$. Then, assuming $a_2 \geq 0$, \eqref{eq:pos_const_2D} is automatically fulfilled, and, if there exists some $a_2$ fulfilling \eqref{eq:mass_const_2D}, feasibility is guaranteed.

\subsection{Control design}
Assuming that the feasibility condition is met, we now extend the reference-governor scheme, which includes the feed-forward strategy, to higher dimensions.  We detail the controller for the leaders in higher dimensions, and, subsequently, we examine the governor loop.

\subsubsection{Leaders control}
The leaders' control steers $\rho^L$ toward a desired time-varying density $\hat{\rho}^L$. Following the method in Section \ref{sec:lead}, we choose
\begin{align}\label{eq:leaders_flux_hd}
\nabla \cdot \left[\rho^L(\mathbf{x}, t)\mathbf{u}(\mathbf{x}, t)\right] = - \hat{\rho}^L_t(\mathbf{x}, t) - K_L e^L(\mathbf{x}, t).
\end{align}
This ensures that
\begin{align}
e^L_t(\mathbf{x}, t) = -K_L e^L(\mathbf{x}, t),
\end{align}
establishing point-wise exponential convergence. To obtain $\mathbf{u}$ explicitly, we add the irrotationality condition
\begin{align}
\nabla \times \left[\rho^L(\mathbf{x}, t)\mathbf{u}(\mathbf{x}, t)\right] = 0,
\end{align}
and solve the resulting Poisson equation using Fourier series, following \cite{maffettone2023hybrid}. The periodicity of $\mathbf{u}$ and conservation of leaders' mass follow from the same argument used in Section \ref{sec:lead}; namely, from the fact that
\begin{align}
\int_\Omega -\left[ \hat{\rho}^L_t(\mathbf{x}, t) + K_L e^L(\mathbf{x}, t)\right] \,\mathrm{d}\mathbf{x} = 0.
\end{align}

\subsubsection{Governor design}
Under the control action \eqref{eq:leaders_flux_hd}, the leaders' density evolves as
\begin{align}
    \rho^L(\mathbf{x}, t) = \hat{\rho}^L(x, t) + {\Lambda}(\mathbf{x}, t),
\end{align}
where $\hat\rho^L$ is the desired time-varying density and ${\Lambda}$ represents the transient behavior:
\begin{align}
    {\Lambda}(\mathbf{x}, t) = -\left[\hat{\rho}^L(\mathbf{x}, 0) + \rho_0^L(\mathbf{x})\right]\mathrm{exp}(-K_L t).
\end{align}
Consequently, $\mathbf{v}^{FL}$ in \eqref{eq:vfl_Rd_def} can be decomposed as
\begin{align}\label{eq:vfl_decomp_hd}
    \mathbf{v}^{FL}(\mathbf{x}, t) = \hat{\mathbf{v}}^{FL}(\mathbf{x}, t) + (\mathbf{f}*{\Lambda})(\mathbf{x}, t),
\end{align}
where $\hat{\mathbf{v}}^{FL} = \mathbf{f}*\hat\rho^L$. Next, we derive an expression for $\hat{\mathbf{v}}^{FL}$ that asymptotically achieves the control goal. This expression will be deconvolved to obtain $\hat\rho^L$ for the leaders to track.

We set the following expression for the velocity field:
\begin{align}\label{eq:vfl_fb_Rd}
{\hat{\mathbf{v}}^{FL} }(\mathbf{x}, t) = \bar{\mathbf{v}}^{FL} (\mathbf{x}) + \alpha(t) \mathbf{w}(\mathbf{x}, t),
\end{align}
where $\bar{\mathbf{v}}^{FL}$ is derived from \eqref{eq:vfl_bar_Rd}. Here, $\alpha {(t)} \in [0, 1]$ is a control function to be determined, and
\begin{align}\label{eq:w_Rd}
\mathbf{w}(\mathbf{x}, t) = \frac{D \nabla\bar{\rho}^F(\mathbf{x})e^F(\mathbf{x}, t)}{\bar{\rho}^F(\mathbf{x})(\bar{\rho}^F(\mathbf{x}) - e^F(\mathbf{x}, t))}.
\end{align}
defines the additional feedback term adjusted by $\alpha(t)$.

Substituting \eqref{eq:vfl_decomp_hd} into \eqref{eq:followes_Rd} (accounting for \eqref{eq:vfl_fb_Rd} and \eqref{eq:w_Rd}) and expressing the equation in terms of the error $e^F$, we derive
\begin{multline}\label{eq:eF_Rd}
    e^F_t(\mathbf{x}, t) = D\nabla^2e^F (\mathbf{x}, t)\\-D\left[1-\mathrm{exp}(-K_Lt)-\alpha(t)\right] \nabla\cdot\left[e^F(\mathbf{x}, t)\frac{\nabla\bar{\rho}^F(\mathbf{x}, t)}{\bar{\rho}^F(\mathbf{x}, t)}\right]\\
    +\mathrm{exp}(-K_L t)\times \\\times\nabla \cdot \left[\left(\bar{\rho}^F(\mathbf{x}, t)-e^F(\mathbf{x}, t)\right) (\mathbf{f}*\rho^L_0)(\mathbf{x})-D\nabla\bar{\rho}^F(\mathbf{x})\right].
\end{multline}
\begin{theorem}\label{th:conv_higher_d}
   Let the control problem be feasible according to Proposition \ref{prop:feasib2D}, and define
\begin{align}
G_1(\mathbf{x}) &= \nabla \cdot \left[\frac{\nabla \bar{\rho}^F(\mathbf{x})}{\bar{\rho}^F(\mathbf{x})}\right].
\end{align} 
If $\Vert G_1 \Vert_\infty < 2$, then \eqref{eq:eF_Rd} converges globally to 0 in $\mathcal{L}^2(\Omega)$.
\end{theorem}
\begin{proof}

    Choosing $\Vert e^F\Vert_2^2$ as a Lyapunov functional for \eqref{eq:eF_Rd}, we obtain
    \begin{multline}
        \left(\Vert e^F(\cdot, t) \Vert_2^2\right)_t = 2D\int_\Omega e^F(\mathbf{x}, t)\nabla^2e^F(\mathbf{x}, t)\,\mathrm{d}\mathbf{x}\\
        \begin{aligned}
            &-2D\left[1-\mathrm{exp}(-K_Lt)-\alpha(t)\right]\times\\&\times\int_\Omega e^F(\mathbf{x}, t) \nabla\cdot\left[e^F(\mathbf{x}, t)\frac{\nabla\bar\rho^F(\mathbf{x})}{\bar\rho^F(\mathbf{x})}\right]\,\mathrm{d}\mathbf{x}+2\mathrm{exp}(-K_L t)\times \\&\times\int_\Omega e^F(\mathbf{x}, t) \nabla \cdot [\left(\bar{\rho}^F(\mathbf{x}, t)-e^F(\mathbf{x}, t)\right) (\mathbf{f}*\rho^L_0)(\mathbf{x})-
        \end{aligned}
        \\D\nabla\bar{\rho}^F(\mathbf{x})]\,\mathrm{d}\mathbf{x}.
    \end{multline}
    Utilizing vectorial identities and the divergence theorem, this can be simplified to
    \begin{multline}
        \left(\Vert e^F(\cdot, t) \Vert_2^2\right)_t = -2D\Vert \nabla e^F(\cdot, t)\Vert_2^2\\
        \begin{aligned}
            &-D\left[1-\mathrm{exp}(-K_Lt)-\alpha(t)\right] \int_\Omega (e^F(\mathbf{x}, t))^2 G_1(\mathbf{x})\,\mathrm{d}\mathbf{x}\\&
        -\mathrm{exp}(-K_Lt)\int_\Omega (e^F(\mathbf{x}, t))^2 H_1(\mathbf{x})\,\mathrm{d}\mathbf{x} 
        \end{aligned}\\
        + 2\mathrm{exp}(-K_Lt) \int_\Omega e^F(\mathbf{x}, t) H_2(\mathbf{x})\,\mathrm{d}\mathbf{x},
    \end{multline}
    where $H_1 = \nabla \cdot (\mathbf{f}*\rho^L_0)$  and $H_2 = \nabla\cdot(\bar{\rho}^F(\mathbf{f}*\rho^L_0) - D\nabla\bar\rho^F)$. By exploiting bounds similar to those derived for Theorem 3 and 4, we establish
    \begin{multline}
        \left(\Vert e^F(\cdot, t) \Vert_2^2\right)_t\leq [-2D+D\Vert G_1(\cdot)\Vert_\infty]\Vert e^F(\cdot, t) \Vert_2^2 \\
        \begin{aligned}
        &+ \Vert H_1(\cdot) \Vert_\infty\mathrm{exp}(-K_Lt)\Vert e^F(\cdot, t) \Vert_2^2\\&+ 2\Vert H_2(\cdot)\Vert_2\mathrm{exp}(-K_L t) \Vert e^F(\cdot, t) \Vert_2.
        \end{aligned}
    \end{multline}
    Under the theorem hypothesis, the bounding system is in the form discussed in Lemma \ref{lemma:bounding_sys}, proving the theorem.
\end{proof}
\begin{figure*}
     \centering
     \begin{subfigure}[b]{0.32\textwidth}
         \centering
         \includegraphics[width=\textwidth]{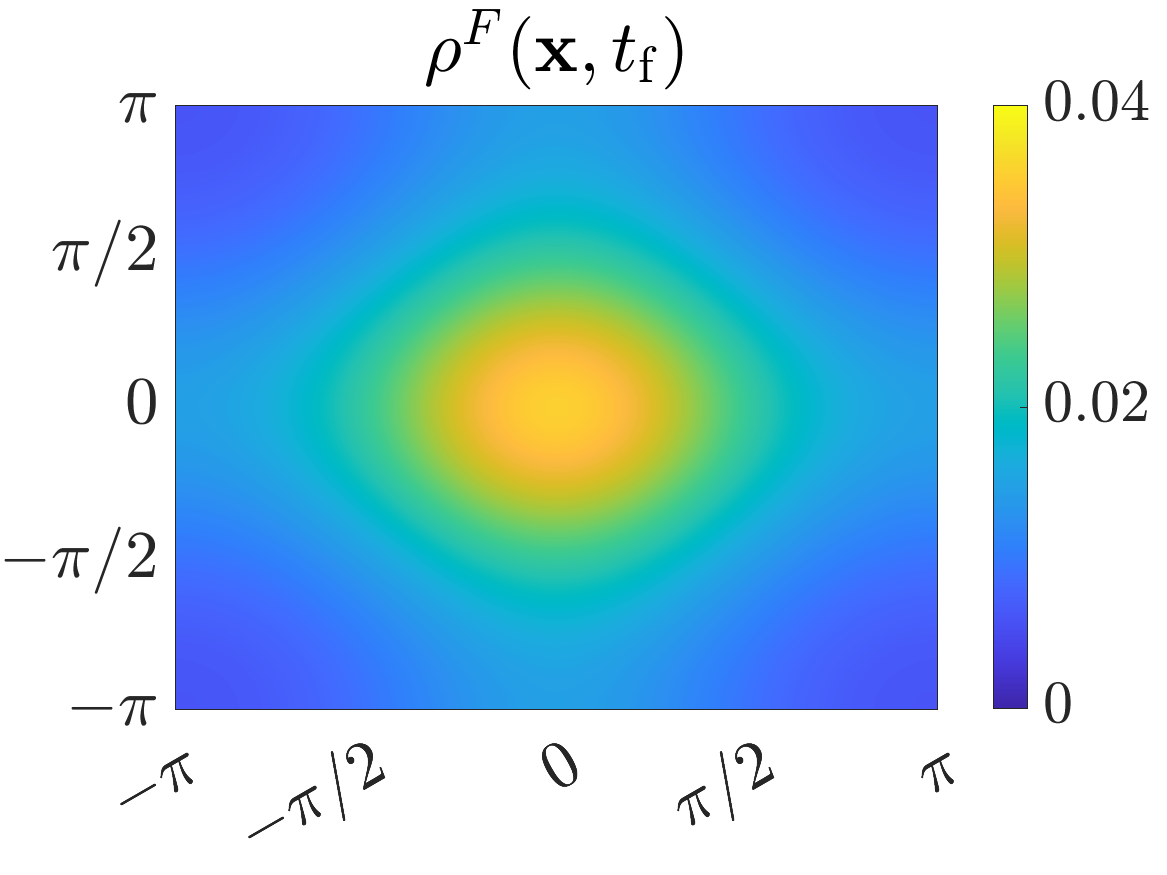}
         \caption{}
         \label{sub:foll_fin_disp}
     \end{subfigure}
     \begin{subfigure}[b]{0.32\textwidth}
         \centering
         \includegraphics[width=\textwidth]{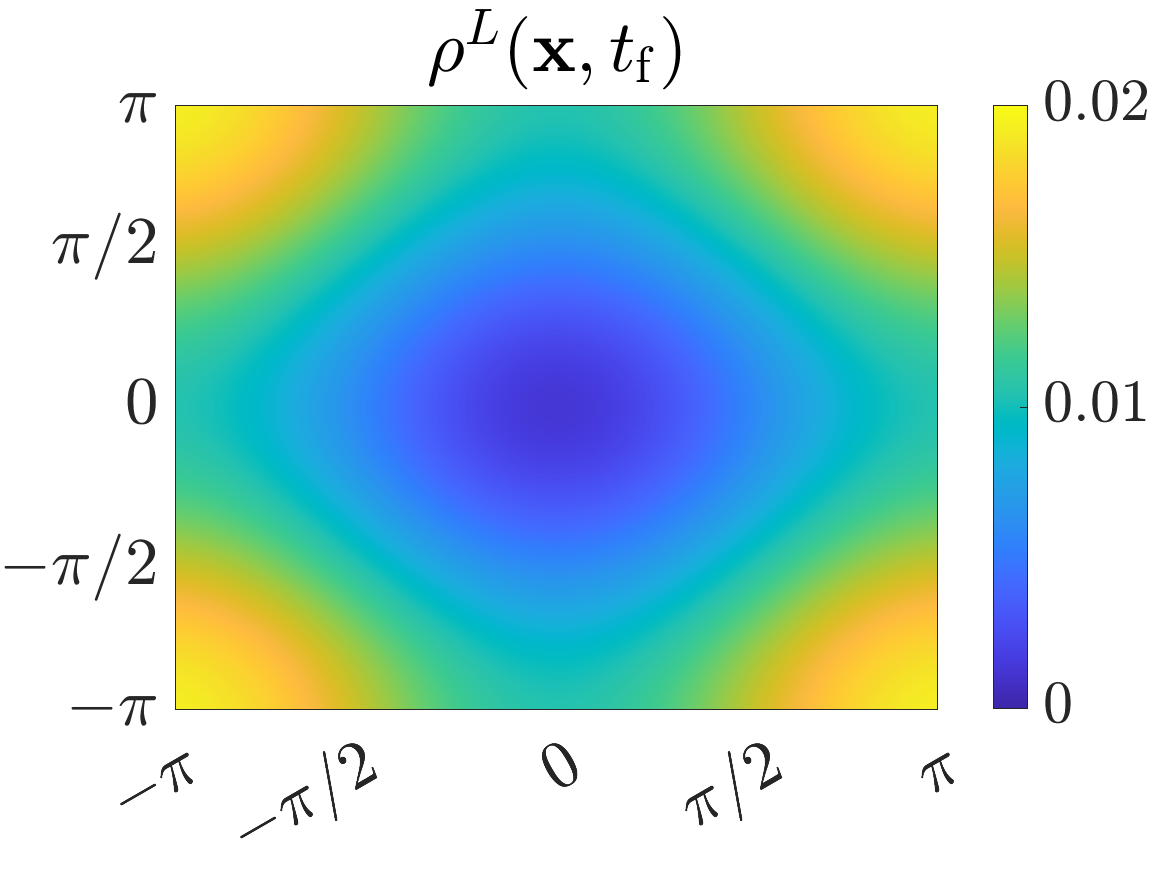}
         \caption{}
         \label{sub:lead_fin_disp}
     \end{subfigure}
     \begin{subfigure}[b]{0.32\textwidth}
         \centering
         \includegraphics[width=\textwidth]{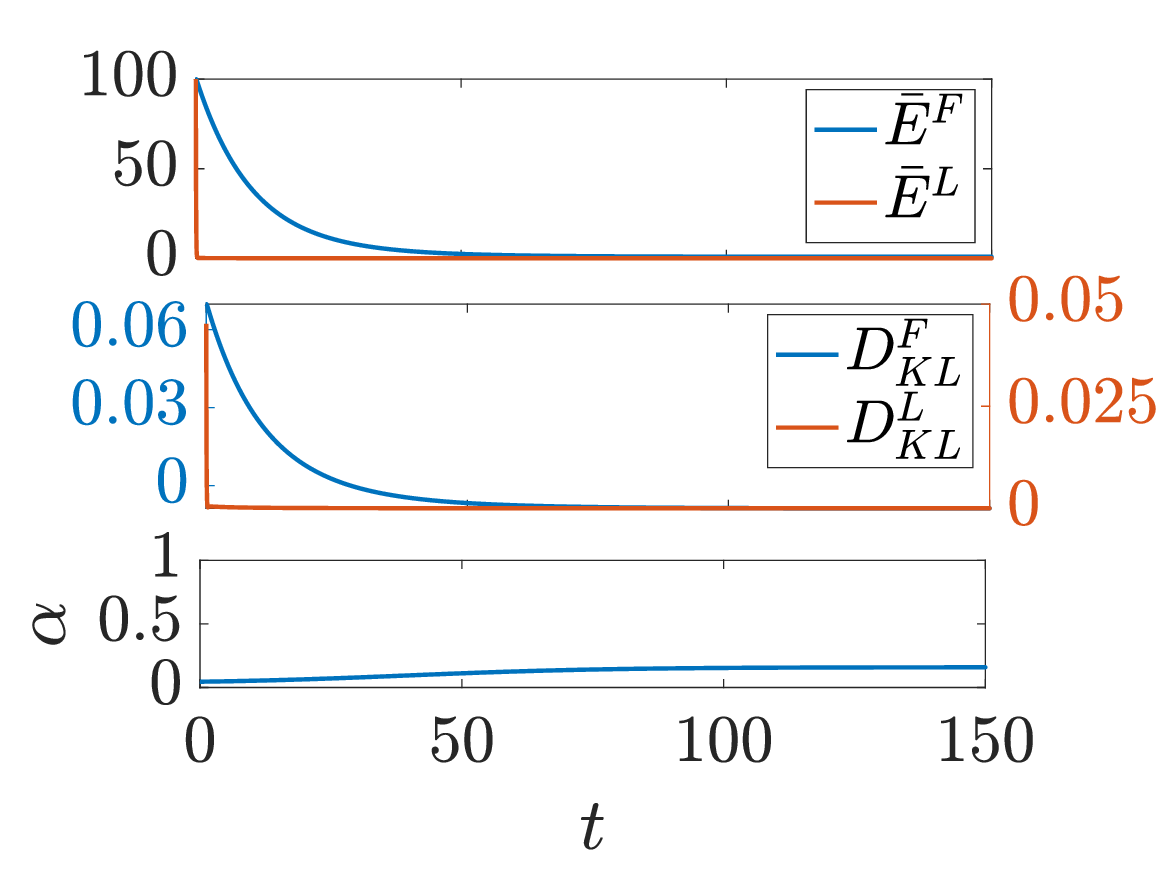}
         \caption{}
         \label{sub:2dtrialres}
     \end{subfigure}
        \caption{\footnotesize Monomodal trial in 2D: (a) followers' density at the end of the trial; (b) leaders' density at the end of the trial; and (c) time evolution of the percentage error (top panel), KL divergences (middle panel), and $\alpha$ (bottom panel).}
        \label{fig:vonMises_2D}
\end{figure*}
 
Convergence is ensured for any $\alpha \in [0, 1]$. The case where $\alpha = 0$ for all $t \geq 0$ coincides with the feed-forward scheme proposed in the one-dimensional case. By performing a deconvolution of $\mathbf{v}^{FL}$, we derive
\begin{align}\label{eq:rho_hat_Rd}
\hat{\rho}^L(\mathbf{x}, t) = \bar{\rho}^L(\mathbf{x}) + \alpha(t) W(\mathbf{x}, t),
\end{align}
where $\bar{\rho}^L$ comes from \eqref{eq:deconv_Rd} and $W$ represents the deconvolution of \eqref{eq:w_Rd}, expressed as:
\begin{align}
W(\mathbf{x}, t) = Q(\mathbf{x}) + \beta(t),
\end{align}
with $\beta(t)$ being an arbitrary time-dependent function.

Note that the deconvolution $W$ is defined up to an arbitrary function of time due to the linearity of the convolution operator and the assumption that the kernel is odd. Consequently, the computation of $Q$ must be performed numerically, as no closed form for the periodic kernel in higher dimensions has been established. Similarly, to the one-dimensional case, $\alpha$ and $\beta$ can be selected such that
\begin{subequations}
\begin{align}
    \int_{\Omega} W(\mathbf{x}, t) \,\mathrm{d}\mathbf{x} = 0, \;\;\forall t\geq 0\\
    \bar{\rho}^L(\mathbf{x}) + \alpha(t) W(\mathbf{x}, t) \geq 0, \;\; \forall t\geq 0.
\end{align}
\end{subequations}
We can choose $\alpha$ using the same rationale in Section \ref{sec:alpha_choice}.

\subsection{Numerical validation} \label{subsec:num_valid2D}
For validation, we extended the trial depicted in Fig. \ref{fig:centered_vonMises} from one to two dimensions. Specifically, we set $D=0.05$, $M^F = 0.6$, and $K_L = 10$. For the desired followers' density, we adopted the two-dimensional version of \eqref{eq:vonmises} -- see Equation (25) in \cite{maffettone2023hybrid} for an explicit formula -- with the concentration coefficients in each direction set at $k_1 = k_2 = 0.5$. This configuration satisfied the feasibility condition.

Using the reference-governor scheme and selecting $\alpha$ as outlined in Section \ref{sec:alpha_choice}, we numerically integrated \eqref{eq:followes_Rd} and \eqref{eq:leaders_Rd} using a central finite difference scheme on a 50$\times$50 mesh. The forward Euler method was employed to estimate time derivatives, with a time step of $\mathrm{d}t = 0.01$. Starting from a constant initial density for both populations, we observed the results shown in Fig. \ref{fig:vonMises_2D}. Both the percentage errors and KL divergences converged to zero within approximately 50 time units, and the weighting factor $\alpha$, which adjusts the amplitude of the feedback correction, stabilized at about 0.2.

\section{Conclusions}
We developed a continuum framework to address the leader-follower density control problem within large-scale multi-agent systems. 
We established criteria for assessing the problem's feasibility, leveraging information about the number of reactive leaders in the group, the desired followers' density, the interaction kernel scale, and the followers' dynamics. Both the proposed control architectures ensure global stability towards a desired spatial organization, for one and multi-dimensional domains. Differently from relevant literature \cite{almi2023, ascione2023mean}, we provided closed forms for the macroscopic control actions and useful bounds for the rate of convergence.

Although convergence is ensured in the limiting scenario of infinite populations, we demonstrated a straightforward methodology to apply our macroscopic control action to swarms of finite size, taking inspiration from \cite{maffettone2022continuification, maffettone2023continuification, maffettone2023hybrid}. 
We emphasize that microscopic analytical guarantees of convergence for swarms of finite size are still missing. Such guarantees could be explored using classical works about two-scale convergence \cite{allaire1992homogenization} and asymptotic formal analysis \cite{sanchez1980non}. 

This is not the only limitation of the study that calls for future research. In fact, future work should aim at ($i$) overcoming the kinematic assumption that is used to model the populations' motion as mass conservation laws; ($ii$) accounting for topological and networked interactions, through, for example, the use of graphons \cite{lovasz2012large}; ($iii$) introducing in the model interactions taking place between followers; ($iv$) analytically study the different robustness properties of the two control schemes we propose, which here where only numerically addressed; and ($v$) proposing an experimental, localized and distributed validation of the strategies within the mixed-reality framework described in \cite{maffettone2023hybrid} -- in so doing, local density estimation methods need to be exploited \cite{izenman1991review}.

Despite these limitations, the proposed work makes contributions to the theory of  density-control of large ensembles that are expected to find application in critical engineering areas such as traffic control and swarm robotics, and opens the door to mathematical treatment of control problems in continuum models describing heterogeneous teams.


\section*{Acknowledgments}
The authors thank Beniamino di Lorenzo (Scuola Superiore Meridionale), for his contributions to the numerical validation in Section \ref{sec:extension}, and the anonymous reviewers whose constructive comments significantly improved the manuscript, particularly in identifying the expression in \eqref{eq:opt_alpha}.

\appendix
\subsection{Kernel periodization} \label{app:kern_periodicization}
Periodic interaction kernels $f$ are obtained from the periodization of standard non-periodic kernels $\hat{f}$,
\begin{equation}
    \label{eq:q_T_def}
    f(x) = \sum_{k=-\infty}^\infty \hat{f}(x+2k\pi).
\end{equation}
\textit{Repulsive kernel}: The non-periodic repulsive kernel is in the form
\begin{equation}
    \hat{f}(x) = \mathrm{sgn}(x) {\mathrm{exp}}\left(-\frac{\lvert x \rvert}{L}\right).
\end{equation}
Note that we utilize a length-scale $L$ while fixing the domain to $[-\pi,\pi]$. Periodization leads to
\begin{equation}
   f(x) =  \sum_{k=-\infty}^{\infty} \mathrm{sgn}(x+2k\pi){\mathrm{exp}}\left({-\frac{\lvert x+2k\pi\rvert}{L}}\right).
\end{equation}
By separating the infinite series into two other infinite series based on the sign of $x+2k\pi$ and computing each of these series individually leads to \eqref{eq:rep_kern}.

\subsection{Deconvolution}\label{app:deconv}
Given the kernel \eqref{eq:rep_kern} and a density function $\rho:\mathcal{S} \rightarrow \mathbb{R}_{\geq 0}$
\begin{multline}\label{eq:123}
	\phi(x) = (f*\rho)(x) = \frac{1}{\mathrm{e}^{\frac{2\pi}{L}} - 1}\bigg[\mathrm{e}^{\frac{2\pi-x}{L}} \int_{-\pi}^x \mathrm{e}^{\frac{y}{L}} \rho(y) \,\mathrm{d}y \\ - \mathrm{e}^{\frac{x}{L}} \int_{-\pi}^x \mathrm{e}^{-\frac{y}{L}} \rho(y) \,\mathrm{d}y \\{-} \mathrm{e}^{\frac{2\pi+x}{L}} {\int_{x}^\pi}\mathrm{e}^{-\frac{y}{L}} \rho(y) \,\mathrm{d}y {+} \mathrm{e}^{-\frac{x}{L}} {\int_{x}^\pi} \mathrm{e}^{\frac{y}{L}} \rho(y) \,\mathrm{d}y\bigg].
\end{multline}
Differentiating twice with respect to the space variable yields
\begin{align}\label{eq:phi_xx}
\phi_{xx}{(x)} = \frac{\phi{(x)}}{L^2} + 2 \rho_x{(x)}.
\end{align}
Thus, by integration, we can retrieve $\rho$ as follows:
\begin{align}\label{eq:deconv_output}
\rho(x) = \frac{1}{2} \int \left(\phi_{xx}(x) - \frac{\phi(x)}{L^2}\right)\,\mathrm{d}x + B,
\end{align}
where $B$ is an arbitrary constant. See the Supplementary Material at \cite{supp_material} for more details.

\section*{References}
\bibliographystyle{IEEEtran}

\end{document}